\documentclass[preprint,times,tighten]{aastex631}

\newcommand{\nc}{\newcommand}

\newcommand{\HII}{H {\sc ii}}
\nc{\msun}{\ensuremath{\mathrm{M}_\odot}}
\nc{\lsun}{\ensuremath{\mathrm{L}_\odot}}
\nc{\thCO}{$^{13}$CO}
\nc{\CeiO}{C$^{18}$O}
\nc{\kms}{\mbox{km~s$^{-1}$}}
\nc{\Kkms}{\mbox{K\,km~s$^{-1}$}}
\nc{\twCO}{$^{12}$CO}

\nc{\cmsq}{\mbox{cm$^{-2}$}}
\nc{\cmcub}{\mbox{cm$^{-3}$}}

\newcommand{\Tex}{$T_{\rm ex}$}

\shorttitle{High mass star formation in G6.55-0.1}
\shortauthors{S. Sen et al.}
\usepackage{graphicx}	
\usepackage{amsmath}	

\usepackage{url}
\usepackage{hyperref}
\hypersetup{
    colorlinks=true,
    citecolor=blue,
    linkcolor=blue,
    filecolor=blue,      
    urlcolor=blue,
    pdftitle={Overleaf Example},
    pdfpagemode=FullScreen,
    }
\usepackage{comment}

\begin{document}
\graphicspath{{./}{figures/}}



\title[High mass star formation in G6.55-0.1]{Kinematics \& Star Formation in the
Hub-Filament System G6.55-0.1}

\correspondingauthor{Saurav Sen}
\email{saurav.sen@tifr.res.in}

\author{Saurav Sen, Bhaswati Mookerjea}
\affiliation{Department of Astronomy \& Astrophysics, Tata Institute of
Fundamental Research,\\ Homi Bhabha Road, Mumbai 400005, India}
\author{Rolf G\"usten, Friedrich Wyrowski}
\affiliation{Max Planck Institut f\"ur Radioastronomie, Auf dem H\"ugel 69, D-53121 Bonn, Germany}
\author{C. H. Ishwara-Chandra} 
\affiliation{National Centre for Radio Astrophysics (NCRA-TIFR), Pune, 411 007, India}


\begin{abstract}
Hub-filament systems (HFSs) being the potential sites of formation of star clusters and high mass stars, provide a test bed for the current theories that attempt to explain star formation globally. It is thus important to study a large
number of HFSs using both intensity and velocity information to constrain these objects better observationally. We present here a study of the hub-filament system associated with G6.55-0.1 using newly obtained observations of radio continuum and $J$=2--1 transition of CO, \thCO, and \CeiO. The radio continuum maps show multiple peaks that coincide with far-infrared dust continuum peaks indicating the presence of more than one young massive stars in the hub of the HFS. We used the velocity information from the \CeiO(2--1) map to (a) show that the source G6.55-0.1 is not physically associated with the SNR W28 and (b) disentangle and identify the velocity components genuinely associated with
G6.55-0.1. Among the velocity-coherent structures identified, the two filaments at 13.8 and 17.3\,\kms\ contribute a total mass accretion rate of ~3000\,\msun\,Myr$^{-1}$ to the hub. Both the filaments also show V-shaped structure, characteristic of gravitational collapse, in their velocity profile at
the location of the hub. Estimated mass per unit length of the segments of the filaments are  smaller  than the critical line masses derived from virial equilibrium considerations. This suggests that while the filaments are not gravitationally collapsing as a whole, the spectra from the hub indicate that the
inner parts are dynamically decoupled and collapsing to form stars.
\end{abstract}




\section{Introduction}

The interstellar medium (ISM) is observed to be filamentary both in its atomic \citep{Heiles1979} and molecular \citep{Schneider1979} phases. This view is firmly established by data obtained with modern observing facilities with ever improving angular resolution and sensitivities \citep{Andre2014,Hacar2018,Arzoumanian2019}.  Observations have also indicated a direct connection  between the filamentary structure of the ISM and the initial conditions for star-formation \citep[see][for recent reviews]{Andre2014, Hacar2022,Pineda2023}. The large-scale Herschel far-infrared dust continuum emission surveys showed that most dense cores and young stellar objects (YSO) are preferentially formed in association to dense filaments in both nearby \citep{Andre2010} as well as in more distant Galactic Plane \citep{Molinari2010} molecular clouds. The low-mass cores found along parsec-scale filaments \citep{Hartmann2001} likely form from  their internal gravitational fragmentation \citep{Schneider1979,Inutsuka1997}. Most of the dense clumps harbouring high-mass stars and clusters are found at the junction of massive filaments, forming the so-called Hub-Filament Systems \citep[HFS;][]{Myers2009,kumar2020}. Filaments provide connection(s) to additional mass reservoir(s) and could funnel large amounts of material towards cores and clumps \citep[e.g.][]{Peretto2013}. Yet, the connection between the existing  star formation models and the new filamentary conditions of the ISM remains under debate, particularly in the case of high-mass stars \citep[e.g.,][]{Motte2018}. The multi-scale dynamics leading to the movement of matter through the filaments to the cores is understood to be a continuous interplay between turbulence and gravity, with the former driving the non-thermal motions from filaments down to small scales, where gravity begins to dominate the dynamics once the regions reach a surface density above a critical value of $\sim 0.1$\,g cm$^{-2}$ \citep{Ohashi2016,Liu2023}.

Currently there exist two families of competing theories of star formation: in one, which has a fair amount of success in explaining low-mass star formation, supersonic turbulence is the one mechanism responsible for defining the mass reservoirs accessible to individual protostars and, as a result, for setting the stellar initial mass function \citep[e.g.,][]{Krumholz2005, Padoan2020}; the other models which provide a more consistent picture of high-mass star formation predict that the  hierarchical gravitational collapse of molecular clouds is what drives their evolution \citep[e.g.,][]{Ballesteros-Paredes2011,VazquezSemadeni2017,Peretto2007}. The HFSs because of their potential as the cradle of low- and high-mass star formation have recently been the targets of many observational studies \citep[][and references therein]{Peretto2013, Liu2023,Zhou2023}. Longitudinal flows along filaments with rates of $\sim 10^{-4}$--10$^{-3}$\,\msun yr$^{-1}$  converging to clusters of stars suggest that such flows, triggered by a hierarchical global collapse are adequate to form stars \citep{Trevino2019}. Stability analyses based on estimates of virial parameters  of the filaments as well as of the embedded clumps reveal that a vast majority of the clouds are self-gravitating but stable on larger scales. On a parsec-scale however the central clumps and/or cores in the same clouds are undergoing gravitational collapse \citep{Mookerjea2023,Peretto2023}.

The far-infrared continuum source IRAS 17577-2320 shows clear signatures of high-mass star formation such as the multi-peaked compact \HII\ region G6.55-0.1 
 \citep{WoodChurchwell1989} and methanol maser at 13.6\,\kms \citep{Walsh1998,WJKim2019}.
Using Herschel continuum images  and a dust emissivity exponent $\beta =2$ \citet{Paradis2014} estimated dust temperatures of 30 and 25\,K respectively, for the central and the surrounding regions. The source IRAS 17577-2320 has also been identified in the literature as one of the star-forming regions near (in projection) the supernova remnant W28. The far-infrared continuum images  revealed a hub-filament-system with the hub coinciding with the IRAS source and four filaments with a total length of 12\,pc and a total mass of 8.5$\times 10^4$\,\msun\ \citep{kumar2020}. Considering the HFSs to be an important piece of tile in the unsolved puzzle of massive star formation we use here newly obtained radio continuum and molecular line ($J$=2--1 transitions of CO and its isotopes) observations to study the nature of the massive protostar and mass accretion in the HFS associated with G6.55-0.1 located at a distance of 3\,kpc \citep{wienen2012}.  The work particularly aims to identify the filaments associated with the HFS as velocity-coherent structures, study mass accretion through the filaments and their stability against gravitational collapse to understand their role in the formation of massive stars in the region.

\section{Observations}

\subsection{Radio continuum Observations and Data Reduction}

We have mapped the low-frequency radio continuum emission at 750 and 1260\,MHz toward G6.55-0.1 using the upgraded Giant Meterwave Radio Telescope \citep[uGMRT;][]{gupta2017}, India.  The GMRT interferometer consists of 30 antennas, each with a diameter of 45\,m, that are arranged in a Y-shaped configuration \citep{Swarup_GMRT}. Of these, 12 antennas are located randomly within a central region of area $1\times1$\,km$^2$, and the remaining 18 antennas are placed along three arms, each with a length of 14\,km. The shortest and longest baselines are 105\,m and 25\,km, respectively. The configuration enables us to map large- and small-scale structures simultaneously.  The observations (Project code: 43\_035) were carried out on 21 and 22 August 2022 in  Band 4 (550-–850 MHz) and Band 5 (1000-–1460 MHz) with the GMRT Wideband Backend (GWB) correlator configured to have a bandwidth of 400 MHz across 4096 channels. The radio source 3C286 was used as the primary flux calibrator and bandpass calibrator, and the sources 1911-201 (at 750 MHz) and 1822-096 (at 1260 MHz) were used as the phase calibrators. Primary calibrators were observed at the beginning and end of the observation for flux and bandpass calibration. The phase calibrators were observed after each scan (30 mins) of the target to calibrate the phase and amplitude variations over the entire observing period. The map was centered at the position $\alpha_{2000} = $18$^h$00$^m$49.9$^s$ and $\delta_{2000}$=-23\arcdeg 20\arcmin 33\arcsec. The angular sizes of the largest structure observable with the uGMRT are 15$\arcmin$ and 9$\arcmin$ at 750 and 1260\,MHz, respectively. Table\,\ref{tab_gmrt} summarizes the details of the uGMRT observations and data.

\begin{deluxetable}{lcc}
 \tablecaption{Details of radio observations with GMRT \label{tab_gmrt}}
\tablehead{
\colhead{Frequency (MHz)} & \colhead{750} & \colhead{1260}}
\startdata
Observation date & 21 Aug 2022 & 22 Aug 2022 \\
On-source time (h) & 5 & 5\\
Bandwidth (MHz) & 400 & 510 \\
Primary beam & 15\arcmin & 9\arcmin \\
Synthesized beam & 4\farcs5$\times$3\farcs7 & 2\farcs7$\times$2\farcs1\\
Position angle (\arcdeg) & -6.54	& 16.2\\
Noise (mJy beam$^{-1}$)    & 0.2	& 0.1 \\   
\enddata
\end{deluxetable}

The data reduction process of flagging, calibration, imaging, and self-calibration was done using the CASA\footnote{\url{https://casa.nrao.edu/index_docs.shtml}} Pipeline-cum-Toolkit for Upgraded GMRT data REduction (CAPTURE\footnote{\url{https://github.com/ruta-k/CAPTURE-CASA6.git}})
continuum imaging pipeline for uGMRT \citep{Ruta_capture}, which utilizes tasks from Common Astronomy Software Applications \citep[CASA][]{McMullin2007}.

The flux density calibration was done using the scale provided by \citet{Perley2017}. After the initial rounds of editing and calibration, we used the multi-term multi-frequency synthesis \citep[MT-MFS; see][]{Rau2011} algorithm in the {\em tclean}
task to account for possible deconvolution errors in wide-band imaging. In the pipeline, five rounds of phase-only self-calibration were performed before making the final image. These maps were then corrected for primary beam gain.

\subsection{Molecular line observations with APEX}

We used the dual-polarization NFLASH230 facility receiver of the APEX telescope\footnote{APEX, the Atacama Pathfinder Experiment is a collaboration between the Max-Planck-Institut für Radioastronomie, Onsala Space Observatory (OSO), and the European Southern Observatory (ESO).} \citep{guesten2006}, connected to digital FFTS backends with a spectral resolution of 61 kHz (project ID M-0110.F-9518C-2022). The wide IF bandwidth of the spectrometer allows simultaneous recording of the $J$=2--1 CO isotopologues $^{12}$CO, \thCO\ and \CeiO. Atmospheric condition during the observations (Nov 11 2022) were stable with precipitable water vapor of 2.5 mm. The half-power beam width of the APEX is 26\farcs2 (at 230 GHz). Data are presented as main-beam temperatures, using a main-beam efficiency $\eta_{\rm mb}$ = 0.80.

Data were acquired in total power on-the-fly mode, sampling spectra on a 10\arcsec\ grid while slewing the telescope in R.A. The observed field, 750\arcsec$\times$750\arcsec\ in size, was centered at  R.A.=18$^{\rm h}$00$^{\rm m}$49.9$^{\rm s}$ Dec=-23\arcdeg 20\arcmin 33\farcs8 (2000). The reference position at R.A. = 17$^{\rm h}$47$^{\rm m}$51.8$^{\rm s}$ Dec = -21\arcdeg 19\arcmin 58\farcs8 (2000) was clean of \twCO(2--1) emission at the level $T_{\rm mb}$ = -0.15\,K. CO line pointing was established on nearby RAFGL1922. For the APEX observations the spectra were box-smoothed to 0.25\,\kms\ spectra resolution and  second order baseline were removed. The reduced spectra have been binned to a spectral resolution of  0.5\,\kms\ and the final CO data cubes with a pixel size of 10\arcsec\ has an rms of 0.25\,K.

\subsection{Archival Infrared and Sub-millimeter Continuum Data}

We have used archival infrared continuum emission maps of the region observed with Spitzer/IRAC, Spitzer/MIPS, Herschel/PACS, Herschel/SPIRE and APEX/Laboca. The 3.6, 4.5, 5.8 and 8.0\,\micron\ maps at $\sim 2$\arcsec\ resolution were observed as part of the GLIMPSE programme \citep{benjamin2003} and the 24\,\micron\ data were obtained as part of the MIPSGAL programme \citep{carey2009,gutermuth2015}, both observed using Spitzer. The far-infrared emission maps at 70, 160, 250, 350, and 500\,\micron\ were observed with beam-sizes of 5\farcs6, 10\farcs7, 17\farcs6, 23\farcs9 and 35\farcs2 respectively, as part of the Hi-GAL key programme of the Herschel \citep{Molinari2010}.

\section{Overview of the W28 Region \&  G6.55-0.1}

\begin{figure*}
    \centering
   \includegraphics[width = 0.8\textwidth]{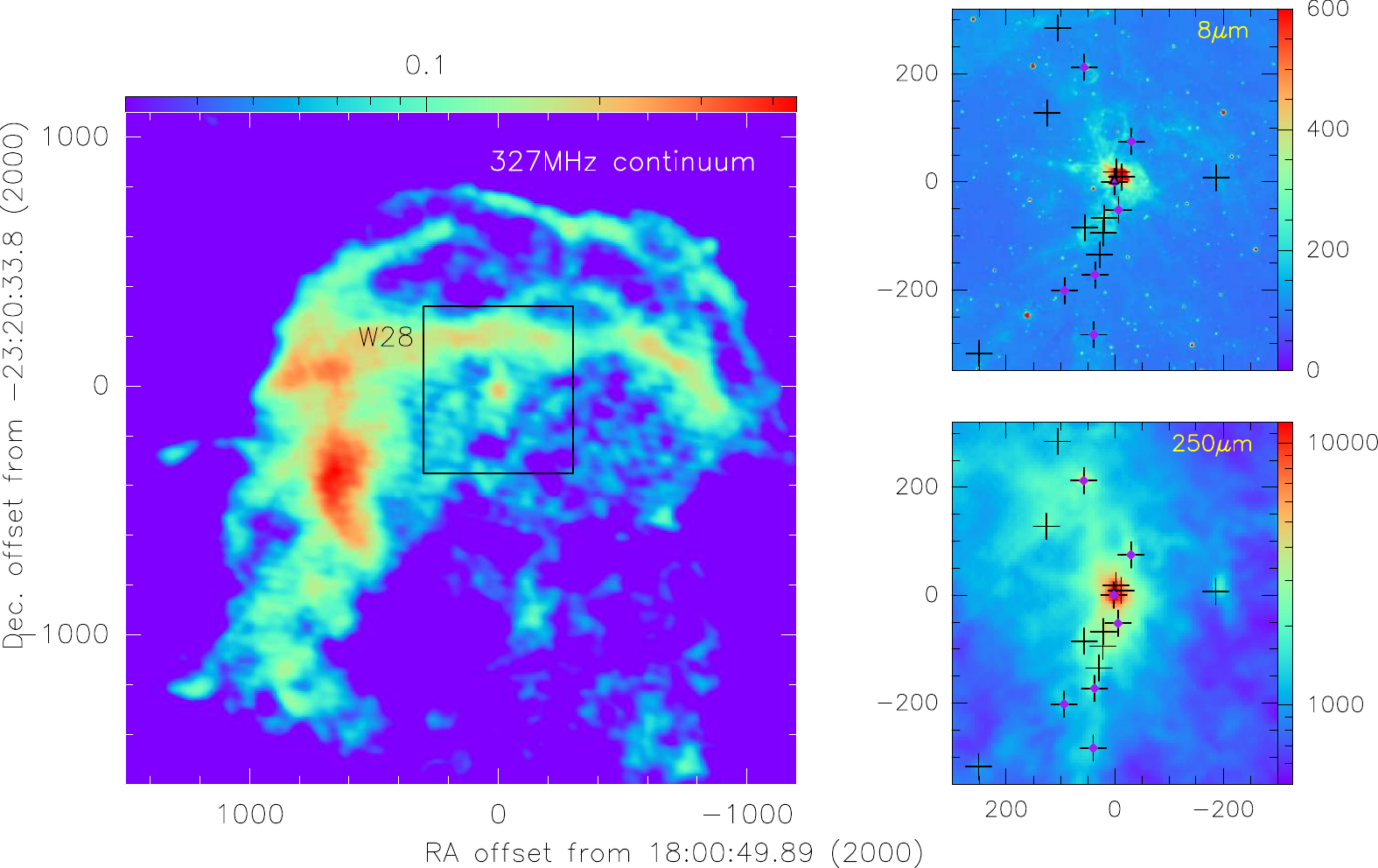}
    \caption{({\em Left} Radio continuum image at 327\,MHz at a resolution of 65\arcsec\ adapted from \citet{Frail1993} with permission. The box shows the region around G6.55-0.1 studied in detail  in this work. ({\em Right}) dust continuum emission from the region G6.55-0.1 at 8\,\micron\ and 250\,\micron\ respectively. Positions of the 17 Hi-GAL continuum sources and 7 GLIMPSE sources are marked with black `+' and purple filled circles respectively. 
    \label{fig_overview}}
\end{figure*}

The radio continuum source G6.55-0.1 is located in the vicinity of the supernova remnant W28 and is associated with a hub-filament-system (HFS) identified in the Herschel/SPIRE 250\,\micron\ image by \citet{kumar2020}. Figure\,\ref{fig_overview} provides an overview of the region in 327\,MHz radio continuum emission \citep{Frail1993} and a close-up view of the emission at 8\,\micron\ arising from the far-ultraviolet irradiated polycyclic aromatic hydrocarbon (PAH) as well the dust continuum emission at 250\,\micron. The radio continuum shows an extended shell  arising from the non-thermal and thermal emission due to the W28 SNR that interacts with the dense molecular clouds in W28F to the east. The region G6.55-0.11 stands out as a bright spot in the otherwise complex emission from W28. The dust continuum emission shows a centrally peaked structure that coincides with the peak of the radio continuum emission but also shows more extended filamentary emission features both to the north and the south. At 250\,\micron\ the filaments are not so clearly discerned in this image, however using dedicated feature-finding algorithms \citet{kumar2020} had identified the source as an HFS. A total of 15 Hi-GAL continuum sources detected in at least two of the five Herschel wavebands have been identified in the region \citep{elia2017} and two additional sources that are detected only in the 70\,\micron\ Herschel image. Figure\,\ref{fig_overview} shows that of these 17 sources nearly 9 lie on the elongated structure extending to the south of the central bright source and seven of these sources have an associated source within 3\arcsec\ in the GLIMPSE catalog. Four of the continuum sources are also detected at 870\,\micron\ in the ATLASGAL images \citep{schuller2009,Csengeri2016}. Table\,\ref{tab_fir} presents the properties of the Hi-GAL sources in the region as estimated by \citet{elia2017} using grey-body fitting of the Herschel flux densities. The derived masses of the continuum sources are typically less than 20\,\msun\ with only three sources including the central source showing masses of 40\,\msun\ or more. Previous studies of CO emission from the region have focussed on the molecular gas affected by the W28 SNR \citep{reach2005,arikawa1999}.  These studies identify the shocked CO emitting gas centred at velocities less than 10\,\kms\ and with CO(2--1) linewidths between 10--30\,\kms \citep{reach2005}.

\begin{deluxetable*}{rcrrrrrrrrrr}
\tabletypesize{\footnotesize} 
\tablecaption{Far-infrared sources in the  G6.55-0.1 region detected by the band-merged Hi-GAL catalogue \citep{elia2017}. The 870\,\micron\ flux is from the ATLASGAL catalogue. \citet{elia2017} estimated dust temperature ($T_{\rm dust}$; Col. 9), mass and luminosity of sources by fitting the SEDs with greybodies. \label{tab_fir}}
\tablehead{
\colhead{Source} & \colhead{Coordinates} & \colhead{$F_{70}$} & \colhead{$F_{160}$} & \colhead{$F_{250}$} & \colhead{$F_{350}$} &
\colhead{$F_{500}$} & \colhead{$F_{870}$} &  \colhead{Mass} & \colhead{$T_{\rm dust}$} & \colhead{$L_{\rm bol}$} & \colhead{D250$^a$} \\
\nocolhead{}& \colhead{(h:m:s d:m:s)} & \colhead{Jy} & \colhead{Jy} & \colhead{Jy} & \colhead{Jy} & \colhead{Jy} & \colhead{Jy}  & \colhead{\msun}  & \colhead{K} & \colhead{\lsun} & \colhead(\arcsec)}
\decimalcolnumbers
\startdata
S1 & 18:01:08.08 -23:25:51.6 & 2.0$\pm$0.1 & 5.2$\pm$0.4 & 5.8$\pm$0.6 & 3.0$\pm$0.4 &\ldots & \ldots & 1.6 & 16.8 & 6.9 & 13.3\\
S2 & 18:00:52.74 -23:25:17.0  & \ldots & 9.5$\pm$2.3 & 25.7$\pm$4.1 & 24.1$\pm$3.0 & 6.0$\pm$0.6 & 0.2$\pm$0.0 & 18.5 & 12.9 & 8.1 & 36.6\\
S3 & 18:00:56.66 -23:23:55.5 & \ldots & 8.7 $\pm$ 1.4 & 7.9$\pm$1.4 & 5.3$\pm$0.3 & 3.1$\pm$0.3 & \ldots & 5.3 & 14.6 & 4.9 & 19.8\\
S4 & 18:00:52.56 -23:23:26.3 & 5.4 $\pm$ 0.2 & 27.8$\pm$1.9 & 43.5$\pm$3.4 & 5.5$\pm$1.5 & \ldots & \ldots & 7.2 & 17.6 & 28.0 & 26.7 \\
S5 & 18:00:51.98 -23:22:48.8 & 3.2$\pm$0.2 & 30.3$\pm$5.2 & 33.7$\pm$3.9 & 8.6$\pm$1.1 & \ldots & \ldots & 5.0 & 19.6 & 25.6 & 24.9\\
S6 & 18:00:51.47 -23:22:08.3 & 3.5$\pm$2.2 & 26.8$\pm$6.2 & 61.2$\pm$39.7 & 13.2$\pm$1.1 & \ldots & \ldots & 11.9 & 15.7 & 56.9 & 32.0 \\
S7 & 18:00:53.95	-23:21:59.1 & 5.2$\pm$2.3 & 5.17$\pm$3.8 & 25.4$\pm$11.4 & 12.5$\pm$13.5 & \ldots  & \ldots & 77.0 & 10.1 & 45.2 & 19.1 \\
S8 & 18:00:51.35	-23:21:41.1  & 73.0$\pm$8.9 & 61.7$\pm$7.7 & 27.3$\pm$21.3 & 70.0$\pm$43.4 & 57.6$\pm$36.1 & \ldots & 9.3 & 19.7 & 162.8 & 13.4\\
S9 & 18:00:49.43	-23:21:25.8 & 4.5$\pm$1.1 & 71.5$\pm$15.3 & 73.1$\pm$40.7 & 12.9$\pm$22.4 &  \ldots &  \ldots & 16.0 & 17.8 & 242.8 & 38.2\\
S10 & 18:00:49.96 -23:20:33.9  & 204.9$\pm$12.9 & 353.7$\pm$18.7 & 314.5$\pm$17.7 & 149.3$\pm$ 4.5 & 71.8$\pm$2.3 & 32.0$\pm$5.1 & 82.7 & 18.7 & 645.1 & 22.6\\
S11 & 	18:00:36.34 -23:20:26.8	 & 4.8$\pm$0.4 & 23.6$\pm$1.8 & 31.6$\pm$1.5 & 13.1$\pm$0.4 & 6.3$\pm$0.3 & 0.6$\pm$0.1 & 11.8 & 15.8 & 25.9 & 23.2\\
S12 & 18:00:49.00 -23:20:25.3\textbf{$^b$} & 340.8  & \ldots & \ldots & \ldots & \ldots & \ldots & \ldots & \ldots & \ldots  \\
S13 & 18:00:49.67 -23:20:15.8\textbf{$^b$}  & 193.7  & \ldots & \ldots & \ldots & \ldots & \ldots & \ldots & \ldots & \ldots  \\
S14 & 18:00:47.68	-23:19:19.0 & 19.5$\pm$ 0.9 & 26.8$\pm$1.8 & 72.5$\pm$6.9 & 23.9 $\pm$ 2.9 & \ldots & \ldots & 40.0 & 13.2 & 56.9 & 23.5\\
S15 & 	18:00:58.99 -23:18:26.6	 & 2.1$\pm$0.2 & 7.7$\pm$0.4 & 11.6$\pm$1.1 & 13.8$\pm$1.2 & 4.9 $\pm$ 0.6 & \ldots & 10.4 & 13.3 & 10.0  & 14.7\\
S16 & 18:00:54.02 -23:17:01.6  & 74.2$\pm$3.8 & 60.5$\pm$2.9& 46.9$\pm$1.9& 17.6$\pm$0.7 & 13.3$\pm$1.3 & 16.3$\pm$2.7 & 4.8 & 23.3 & 166.4 & 11.7\\
S17 & 18:00:57.52 -23:15:49.1  & 4.3$\pm$0.5& 9.0$\pm$1.6 & 15.0$\pm$1.8 & 3.1$\pm$0.5 & \ldots & \ldots & 2.6 & 17.4 & 14.4 & 30.7\\
\enddata
\tablecomments{$^a$ $D250$ is the 250\,\micron\ FWHM. At the distance of 3\,kpc 1\arcsec\ corresponds to 0.015\,pc  $^b$ Source not present in the band-merged catalog \citep{elia2017} but present in the 70\micron\ Hi-GAL catalog \citep{Molinari_2016}.}
\end{deluxetable*}

\section{The ionized gas in G6.55-0.1}

\begin{figure*}
\centering
\includegraphics[width=\textwidth]{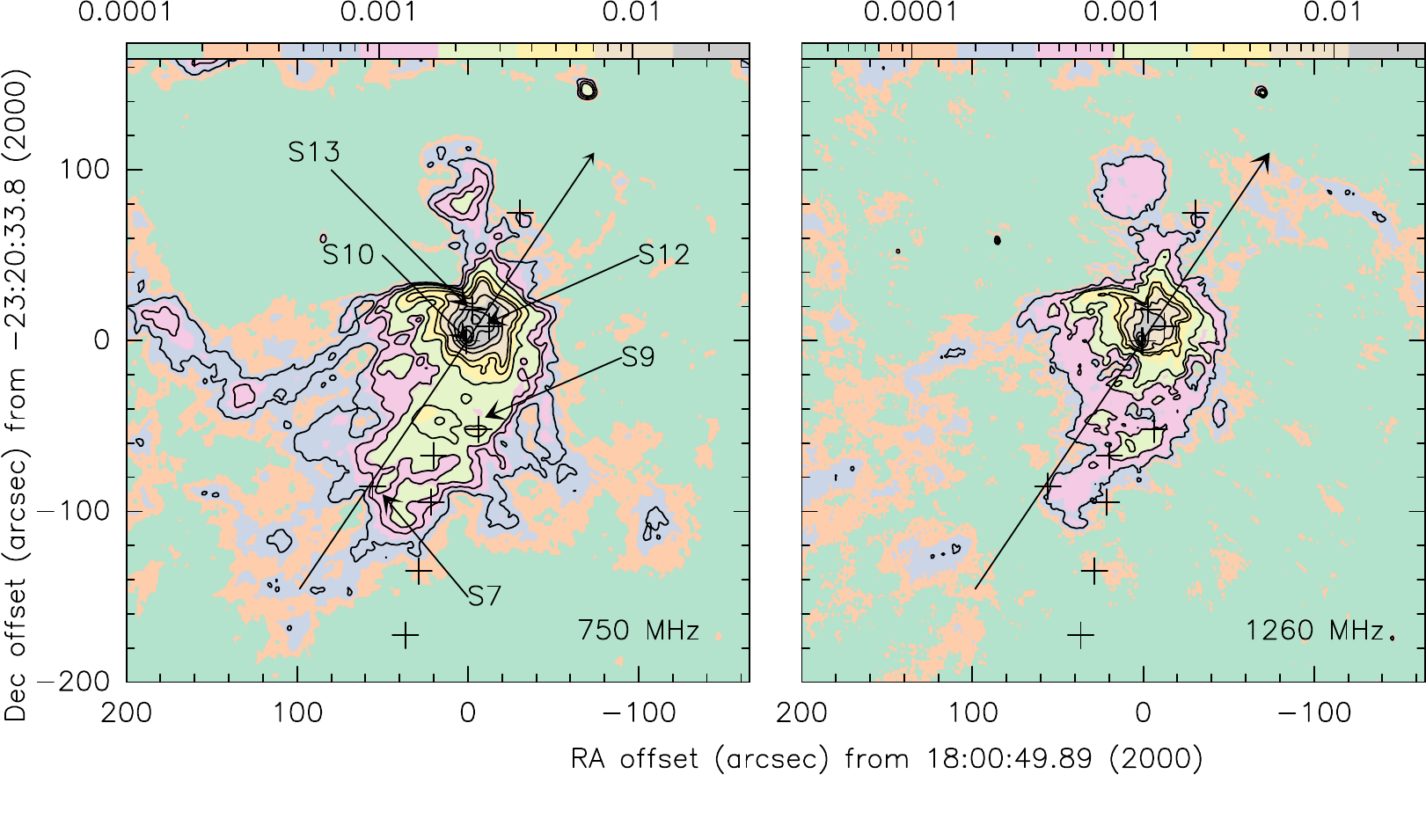}
        \caption{Maps of radio continuum emission at 750 ({\em Left}) and 1260\,MHz ({\em Right}). The coordinates are shown as offsets relative to the center RA: 18$^{\rm h}$00$^{\rm m}$49.89$^{\rm s}$ Dec: -23\arcdeg20\arcmin33\farcs8. The black contours are drawn starting from 3-$\sigma$, with levels (in mJy/beam) at 750\,MHz being 0.6, 1.1, 1.6, 3, 5, 6.5 to 29 (in s.pdf of 0.5) and at 1260\,MHz being 0.3, 0.8, 1.3, 3, 4, 6.5 to 26.5 (in s.pdf of 5). The synthesized beam sizes at 750 and 1260\,MHz are 4\farcs5$\times$3\farcs7 and 2\farcs7$\times$2\farcs1, respectively. The Hi-GAL sources in the region are shown as `+'
      \label{fig_radiocont}}
\end{figure*}

\begin{figure}
    \centering
    \includegraphics[width = 0.5\textwidth]{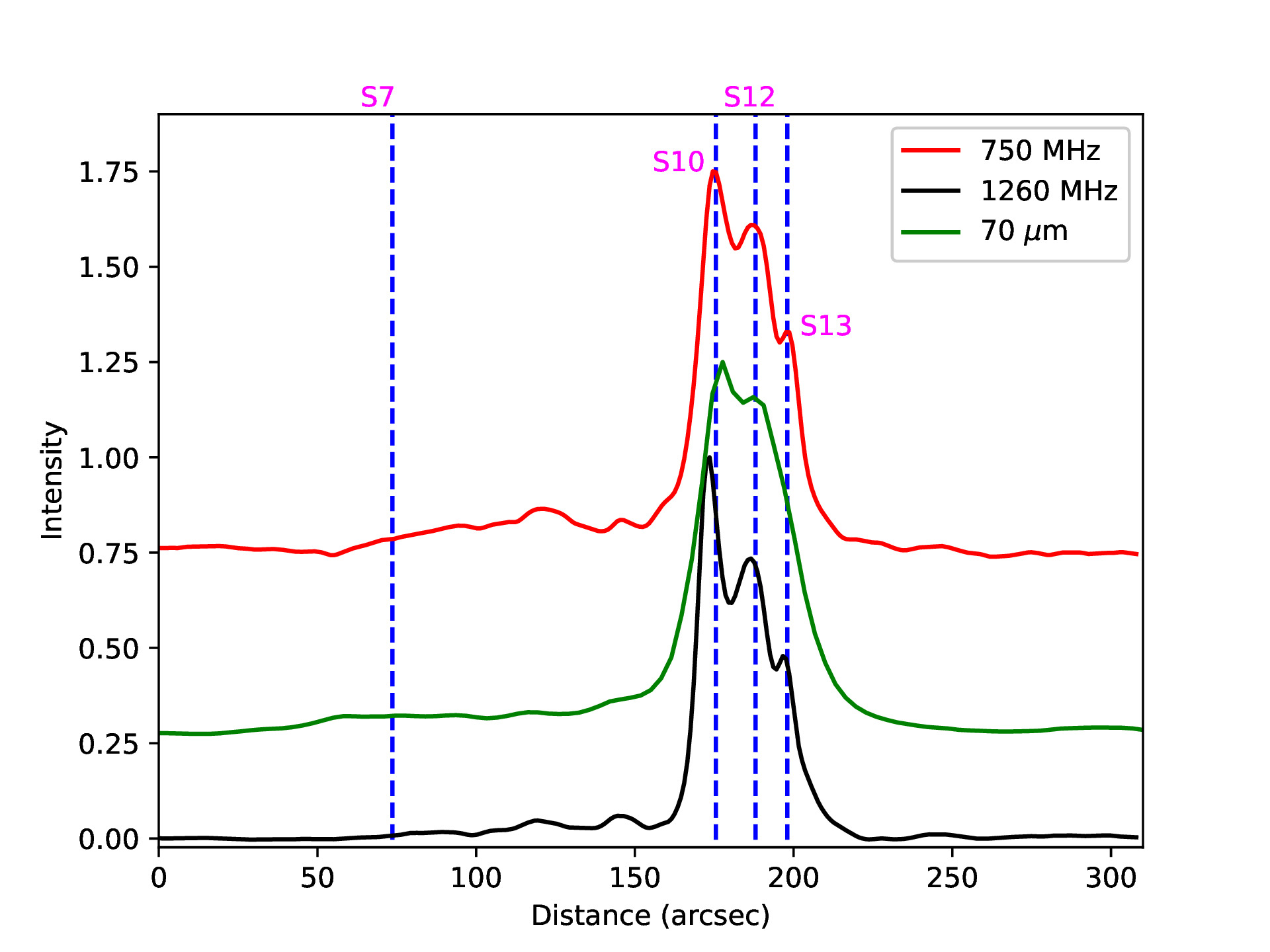}
    \caption{Comparison of the normalised radial profiles as function of the offset from the south-east end of a cut shown in (Fig.\ref{fig_radiocont}) at 70\,\micron, 750 and 1260 MHz. The vertical dashed lines show the positions of the three Hi-GAL sources along the cut. For better visibility the 70\,\micron\ and 750\,MHz profiles are shifted vertically by adding 0.25 and 0.75 respectively. 
    \label{fig_intprof}}
\end{figure}

Figure\,\ref{fig_radiocont} shows the radio continuum images at 750 and 1260\,MHz observed using the uGMRT. At both frequencies the emission shows a compact peak at the position of the \HII\ region G6.55-0.1 with very low intensity extended emission. The emission from the central region though compact show two more local peaks at the high resolution of the radio observations. In order to investigate whether there are any dust continuum peaks co-located with these radio continuum peaks we have derived the intensity profile of the 750\,MHz, 1260\,MHz and the 70\,\micron\ PACS emission along the direction shown in Fig.\,\ref{fig_radiocont}. The radio continuum intensity profiles show three clear enhancements while the 70\,\micron\ profile shows only the main and the secondary peak (Fig.\,\ref{fig_intprof}) that match with the radio peaks. The vertical lines in Fig.\,\ref{fig_intprof} show that the three intensity peaks in the radio continuum coincide with the location of the main source (S10) and two other Hi-GAL sources, and two other sources (S12 and S13) which are detected at 70\,\micron\ only \citep{Molinari_2016}.

We have estimated the radius of the \HII\ region to be 24\arcsec\ by considering the area enclosed by the contour corresponding to the radio continuum intensity of 20\% of the peak value.  The integrated intensity within this region is 1.5 and 2.0 Jy at 750 and 1260\,MHz, respectively. Assuming that the diffuse emission at 1260\,MHz is optically thin, we estimate the Lyman continuum photon rate ($N_{\rm Lyc}$) and the spectral type of the star that causes the ionized emission using the following equation \citep{Mezger1974}

\begin{equation}
\rm{\left[\frac{N_{Lyc}}{s^{-1}}\right]=4.771\times10^{42}\left[\frac{S_\nu}{Jy}\right]\left[\frac{\nu}{GHz}\right]^{0.1}\left[\frac{T_e}{K}\right]^{-0.45}\left[\frac{d}{pc}\right]^2}
,\end{equation}

where $\rm{S_\nu}$ is the flux density at frequency $\nu,$ which is 2.0\,Jy at 1260\,MHz, $T_{\rm e}$ is the electron temperature assumed to be 10$^4$\,K and $d$ is the distance to the source, which is 3\,kpc \citep{Urquhart2018}. The $N_{\rm Lyc}$ is  $1.4\times10^{48}$~s$^{-1}$ and  if it is assumed to be due to a single main-sequence star, then the spectral type of the Zero Age Main Sequence (ZAMS) star is O8.5V with a bolometric luminosity of 5.3$\times 10^4$\,\lsun\ \citep{thompson1984}.
The $L_{\rm bol}$  estimated from the dust continuum flux densities at 70, 160, 250, 350 and 500\,\micron\ over the same region as the \HII\ region is 4.9$\times 10^4$\,\lsun. For this estimate 
we have fitted a two component greybody function to the dust continuum spectral energy distribution assuming dust emissivity exponents of 1 and 2 respectively, for the warm and the cold dust. However, we also note that based on the radio continuum emission distribution the massive stellar component creating the \HII\ region is likely to be more than one early-type star.

A power-law fit to the observed intensities at 750 and 1260\,MHz results in an index of 0.56. The positive spectral index is suggestive of the emission being primarily thermal and arising due to the free-free emission from the \HII\ region created by the massive star(s) therein \citep{Olnon_thermal_emission}. At an angular resolution of $>$1\arcmin\ \citet{Dubner2000} derived an index of -0.35 for the entire W28 region, which likely corresponds to the extended non-thermal emission from the W28 SNR.  However, the positive spectral index also suggests that the emission even at 1260\,MHz is unlikely to be optically thin, which could lead to an underestimate of $N_{\rm Lyc}$.

\section{Molecular material in G6.55-0.1}

\subsection{Column Density from dust continuum  emission \label{sec_grey}}

\begin{figure}
    \centering
    \includegraphics[width = 0.5\textwidth]{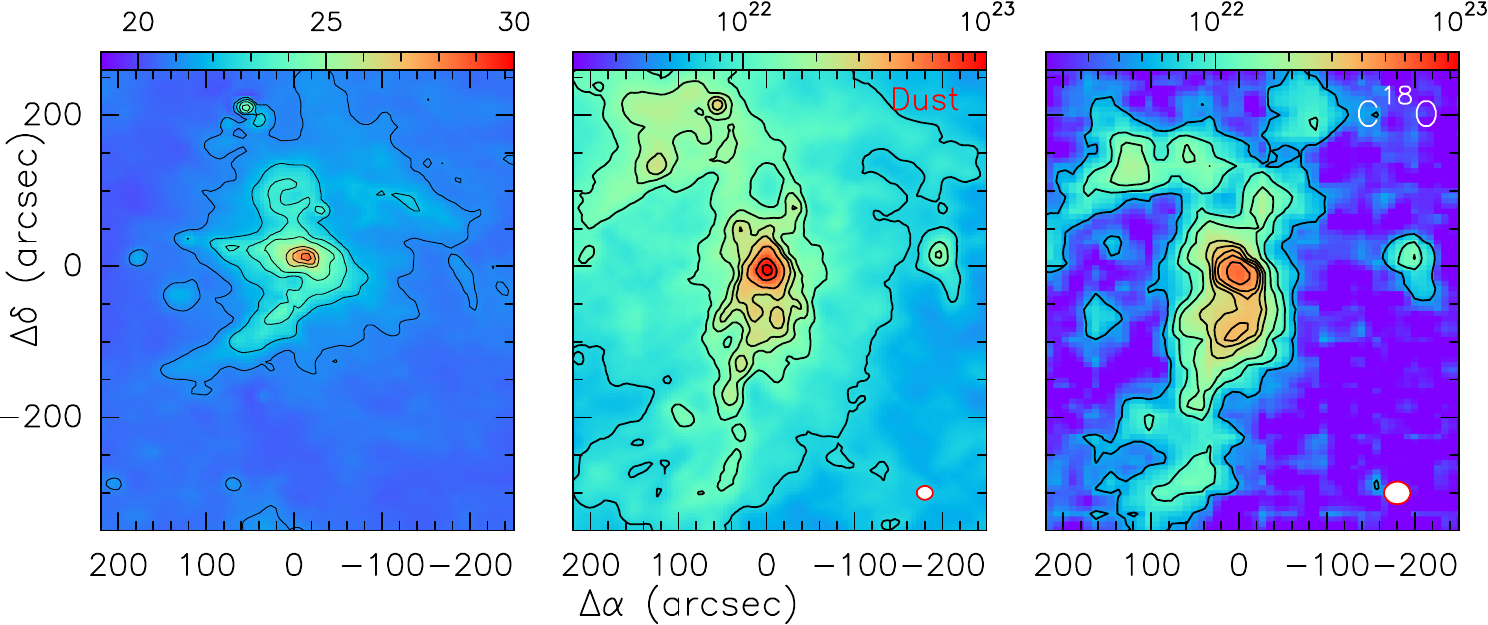}
    \caption{{\em Left:} Dust temperature and column density maps of the region derived from ({\em Middle}) dust continuum at 18\farcs2 resolution created by pixel-by-pixel greybody fitting and removing the diffuse emission following a procedure outlined by \citet{palmeirim2013herschel}. Contour levels are at  0.2--10$\times 10^{22}$ in s.pdf of 0.5$\times 10^{22}$\,\cmsq. ({\em Right}) Column density map from the  \CeiO(2--1) emission intensity integrated between 11 to 19\,\kms\ with a beamsize of 30\arcsec, for $T_{\rm ex}$ = 20\,K. Contour levels are  0.5--6$\times 10^{22}$ in s.pdf of by 0.4$\times 10^{22}$\,\cmsq. The axes are offsets are with respect to R.A. = 18$^{\rm h}$00$^{\rm m}$49.9$^{\rm s}$, Dec=-23\arcdeg 20\arcmin 33\farcs8 (2000).
   \label{fig_colden}}
\end{figure}

In order to obtain an overview of the distribution of cold molecular gas that is essentially the reservoir for material forming the stars, we  used the far-infrared emission maps between 160 to 500\,\micron\ to obtain the distribution of column density of the cold dust in the region. The 70\,\micron\ data was not included in this analysis because at 70\,\micron\ there is still some contribution from smaller dust grains which do not emit much at longer wavelengths and also the presence of hot dust necessitates the use of at least two dust temperature components thus making the fit less constrained. The Herschel images at the four wavelengths were first corrected for the zero-level offsets. For this we assume that the images probe the optically thin dust emission following the empirical fit (e.g., Planck Collaboration XI 2014),

\begin{equation}
I_\nu = \tau_{\nu_0} B_\nu(T)\left(\frac{\nu}{\nu_0}\right)^\beta   
\end{equation}

where $\nu_0$ is the reference frequency at which the optical depth $\tau_{\nu_0}$ is estimated, $I_\nu$ is the specific intensity, and $B_\nu$(T) is the Planck function for the emission of dust at temperature $T$ and frequency $\nu$. All-sky maps of the dust optical depth at $\nu_0$ = 353 GHz (850\,\micron) and the dust temperature maps  generated by the Planck Collaboration XI (2014) at a resolution of 5\arcmin\ were used. For $\beta = 2$, adopting $\nu_0$ = 353\,GHz as the reference frequency and substituting the optical depth at 353\,GHz in Eq. (2) we used the Planck dust temperature map to derive the images of G6.55-0.1 at 160, 250, 350, and 500\,\micron, corresponding to the Herschel wavelengths. By convolving each Herschel image to the Planck resolution, we obtained the offsets of each Herschel image by a comparison with the derived intensity maps \citep{Bernard2010}. The offsets  were found to be 200, 148, 75, and 20\,MJy\, sr$^{-1}$ at 160, 250, 350, and 500\,\micron, respectively. Dust temperature and column density maps of the region were derived by performing pixel-by-pixel greybody fitting of the offset-corrected continuum fluxes using

\begin{equation}
    F_\nu = \Omega B_\nu(T_d)(1-e^{-\tau_{\nu}})
\end{equation}
where, 
\[\tau_{\nu} = \mu m_H \kappa_\nu N ({\rm H_2})s\]
where $F_\nu$ is the flux density measured in a pixel, $\Omega$ is the angular area of each pixel, $B_\nu({\rm T_d})$ is the blackbody function at dust temperature $T_{\rm d}$, $\mu = 2.86$ is the mean molecular weight of H$_2$, $m_{\rm H}$ is the mass of a hydrogen atom, absorption coefficient is given by $\kappa_\nu = 0.1 (\nu/1000 GHz)^\beta$ cm$^2$\,g$^{-1}$, $\beta = 2$ is the dust emissivity index and $\tau_\nu$ is the optical depth. We apply the technique described in \citet{palmeirim2013herschel} that considers  flux measurements up to 500\,\micron\ but  uses a multi-resolution decomposition to generate the column density map at a resolution of 18\farcs2 corresponding to the Herschel beam at 250\,\micron. The fitted dust temperature lies between 20--28\,K and the column density ranges between 5$\times 10^{21}$ to 10$^{23}$\,\cmsq\ (Fig.\,\ref{fig_colden}), with the hub of the HFS showing the maximum column density. The dust temperature is maximum at the location of the \HII\ region and does not show any enhancement towards the north-east where the SNR W28 lies. This suggests that the dust is primarily heated by the embedded high mass star forming region.

\subsection{Molecular line emission \& column density estimates}

\begin{figure*}
    \centering
    \includegraphics[width = 0.9\textwidth]{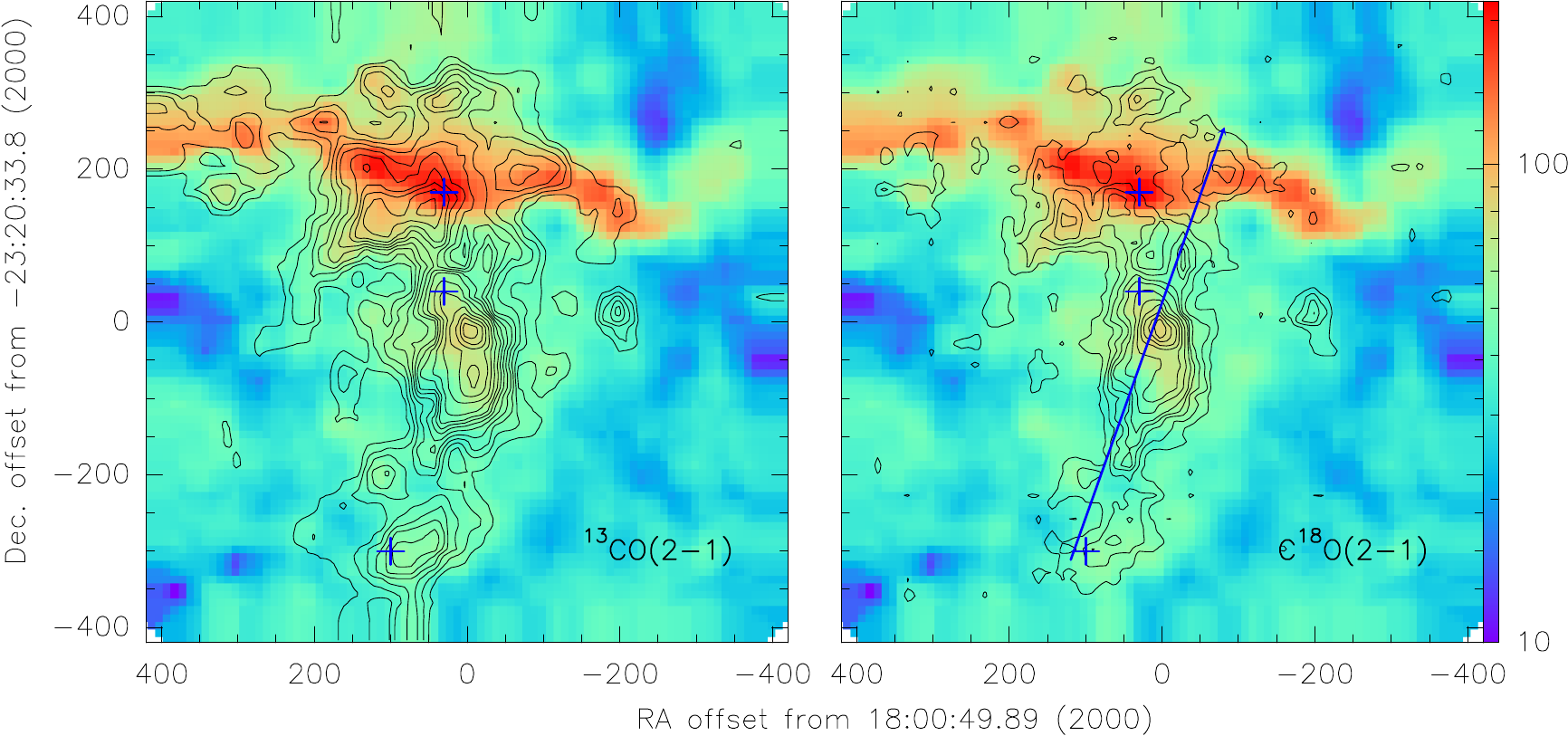}
    \caption{Integrated intensity maps of $J$=2--1 transitions of CO (color) with contours of  ({\em Left}) \thCO\ and ({\em Right}) \CeiO. The CO map is integrated between 4 to 25\,\kms\ and the \thCO\ and \CeiO\ is integrated by 6 to 25\,\kms. The contour levels (in units of K\,\kms) of the \thCO(2--1) map are 7 to 45 (in s.pdf of 2) and of the \CeiO(2--1) map are  1.8 to 15.4 (in s.pdf of 0.8). The `+' shows the selected positions for which the spectra are shown in Fig.\,\ref{fig_cospec}. The blue line with an arrowhead shows the direction along which position-velocity diagrams for the CO, \thCO\ and \CeiO\ emission were studied.}
    \label{fig_coint}
\end{figure*}

\begin{figure*}
    \centering
    \includegraphics[width = 0.85\textwidth]{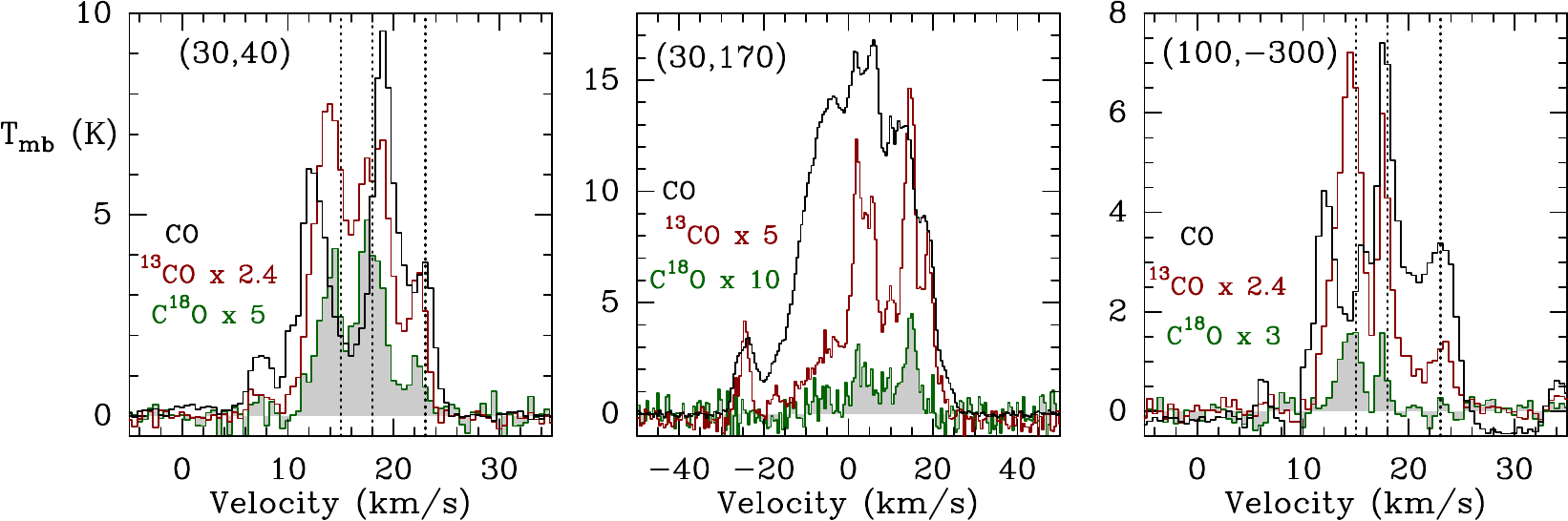}
    \caption{Comparison of CO (black), \thCO\ (red) and \CeiO\ (filled histogram with green outline) spectra at selected positions in the mapped regions. Each spectrum is centered at an offset relative to the center of the map at 18:00:49.93 -23:20:33.89 (2000) marked in the panel and averaged over 30\arcsec. For better visibility the \thCO\ and \CeiO\ spectra are multiplied by factors mentioned in each panel.
    \label{fig_cospec}}
\end{figure*}

The CO(2--1) emission from the mapped region is primarily dominated by the molecular gas associated and most likely interacting with the W28 SNR and is much brighter than the emission from the source G6.55-0.1 (Fig.\,\ref{fig_coint}). In contrast the \thCO\ and in particular the \CeiO\ emission is dominated by  molecular gas spatially overlapping with the \HII\ region G6.55-0.1 and tracing the dust continuum features seen in the 250\,\micron\ map in Fig.\,\ref{fig_overview}. This indicates that the higher intensity CO(2--1) emission in the map corresponds to lower column density high temperature gas coinciding with the W28 shell, whereas the isotopes mainly trace the high column density gas in G6.55-0.1 and trace the northern feature only in a limited region. The CO, \thCO\ and \CeiO\ spectral profiles in the mapped region are complex due to a combination of multiple velocity components as well as optical depth effects particularly for CO and \thCO. We compare the spectra of CO and its isotopes at three positions (marked in Fig.\,\ref{fig_coint}) selected to demonstrate the diversity of velocity profiles detected in the region (Fig.\ref{fig_cospec}). We note that except for the CO spectrum at (30, 170) the spectra at the other two positions do not show any line broadening that might be suggestive of shock heating by the W28 SNR. The \CeiO(2--1) spectra at the `quiescent" positions in the region show three distinct velocity peaks approximately at 15, 18 and 23\,\kms.  While the 23\,\kms\ feature is clearly detected also in the \thCO\ and CO, between 10 and 20\,\kms\ the spectra for both these species are affected by moderate to severe self absorption. The two positions (30, 40) and (100, -300) are representative and capture the basic features of the three spectral features arising from G6.55-0.1 at most positions lying on the north-south extended emission feature that approximately terminates a little to the north of the radio continuum peak. We note that at these two positions the \thCO(2--1) emission also does not arise from gas less than 10\,\kms. We use only the optically thin \CeiO(2--1) spectra for analyses of the velocity components and as well as for the derivation of other physical quantities. 

\citet{wienen2012} had observed NH$_3$ lines at the position of S10, also an ATLASGAL source and identified two components at 12.8 and 16.1\,\kms\ both with a linewidth of 2\,\kms. This suggests that the 18\,\kms\ cloud may not correspond to high density gas. Based on the \CeiO(2--1) spectra (Fig.\,\ref{fig_cospec}) and the channel map (Fig.\,\ref{fig_ceiochanmap}) it is difficult to separate out the extents of the emission from the different components since these spatially overlap. Thus we have used an analysis involving decomposition of the spectrum at each pixel into separate velocity components with Gaussian profiles, followed by the grouping of the velocity coherent structures using a friend-of-friend algorithm (Sec.\,\ref{sec_gausscomp}). 

Considering the \CeiO(2--1) emission seen in the channel map (Fig.\,\ref{fig_ceiochanmap}) as well as the individual spectra (Fig.\,\ref{fig_cospec}) we  estimate the column density of the region around G6.55-0.1 from the optically thin \CeiO\ intensities integrated between 11 to 19\,\kms\ using the following equation: 


\begin{equation}
N({\rm C^{18}O}) = \frac{3h}{8\pi^3 \mu^2}\frac{Z}{\left[J_\nu(T_{ex})-J_\nu(T_{bg})\right]\left(1-e^{\frac{h\nu}{k_B T_{ex}}}\right)}\displaystyle\int {T_{mb}dv}
\end{equation}

with the $Z$ is the partition function given by
\begin{equation}
Z = \frac{k_B T_{ex}}{J_u B h}\exp\left(\frac{B J_u(J_u+1)h}{k_B T_{ex}}\right)  
\end{equation}

where B=5.4891420$\times  10^{10}$\,s$^{-1}$ is the rotational constant for
\CeiO, $\mu$ = 0.11079 D is its dipole moment, and $J_u$ is the upper
level of the transition, equal to 2. The CO and \thCO(2--1) though optically thick are heavily affected by self-absorption so that estimate of $T_{\rm ex}$ is difficult using their spectra. We assume $T_{\rm ex}$ to be 20\,K, similar to the average dust temperature in the region and estimate a factor of 8.8$\times 10^{14}$\,cm$^{-2}$\,K$^{-1}$\,[\kms]$^{-1}$ for conversion of the integrated \CeiO(2--1) intensities to $N$(CO). This factor increases by a factor of 2 for an assumed $T_{\rm ex}$ of 10\,K. We further adopt $^{16}$O/$^{18}$O to be 500 and CO/H$_2$=10$^{-4}$ to estimate the molecular H$_2$ column density, $N$(H$_2$), for the entire region (Fig.\,\ref{fig_colden}). The $N$(H$_2$) distribution so obtained from the velocity-selected \CeiO(2--1) emission is similar to the distribution derived from dust continuum emission, this implies that the molecular line indeed traces the HFS detected in the 250\,\micron\ image by \citet{kumar2020}. We note that in the \CeiO\ column density map an extra emission feature is detected towards the north-west of the hub, which is not seen in dust emission. The column density estimate from the \CeiO\ emission ranges between (0.5--6)$\times 10^{22}$\,\cmsq\ which is less than the column density estimated from the dust emission by 60\%. However considering the fact that the \CeiO(2--1) intensity is for the velocity range 11--19\,\kms\ identified to be associated with G6.55-0.1, while the dust column density includes all components along the line of sight and also the uncertainty in the conversion factor due to the assumed \Tex, the column density estimates from dust and gas are consistent with each other. Based on the column density distribution we estimate the HFS G6.55-0.1 to be 7.6\, pc long with a width of 2.3\,pc at most places, with a total mass of 4520\,\msun. The mass estimated for the HFS is much smaller than the value of 8.5$\times 10^4$\,\msun\ estimated earlier for a larger structure  with four filaments as analyzed by \citet{kumar2020}.

\subsection{Velocity distribution in the region}

\begin{figure*}
    \centering
    \includegraphics[width = 0.8\textwidth]{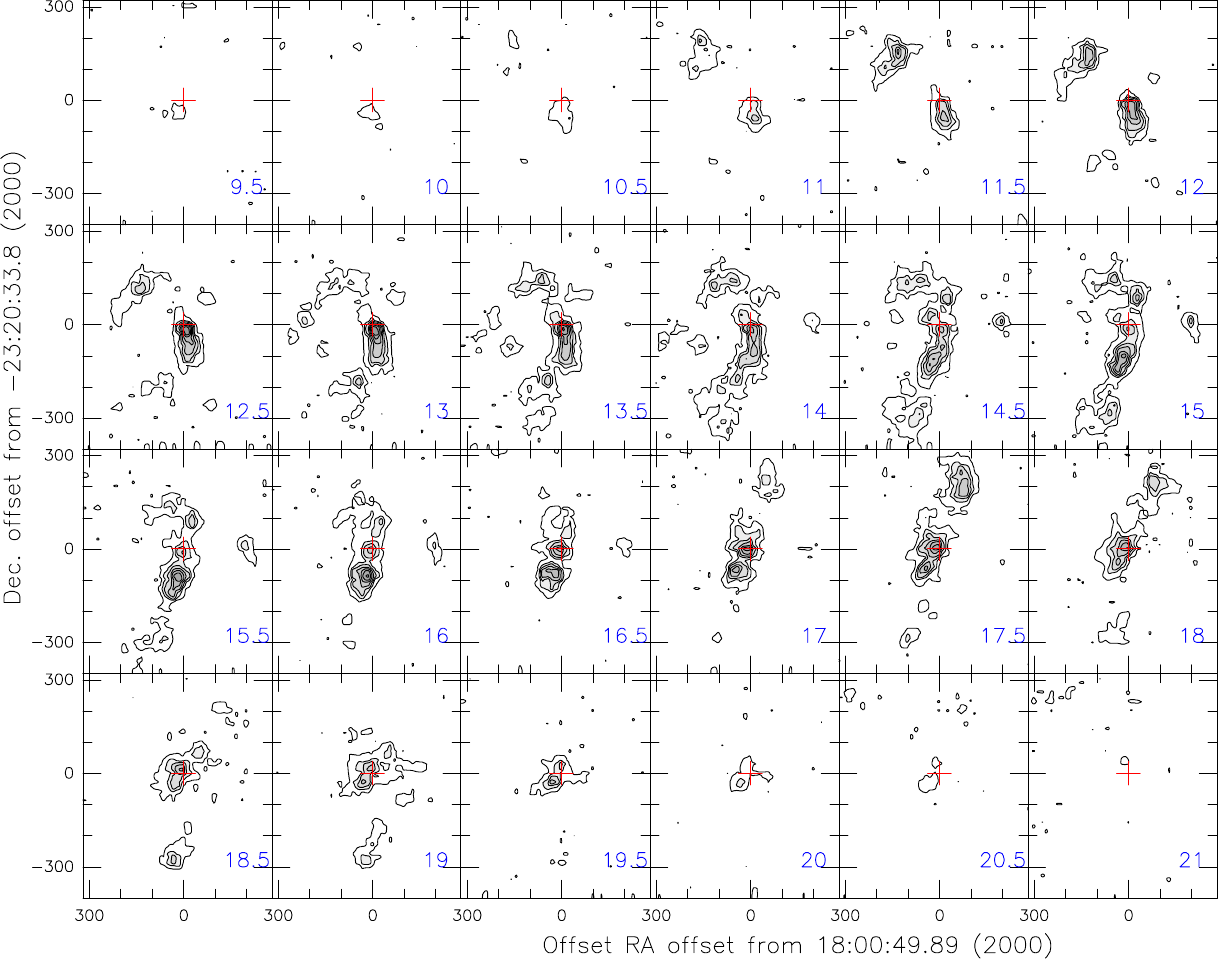}
    \caption{Velocity-channel map of the \CeiO(2--1) emission. The contour levels are at  0.3, 0.7, 1, 1.4, 1.8, 2.2, 2.6 and 3\,\kms. The nominal position of the center of the map is marked with a red `+'. 
    \label{fig_ceiochanmap}}
\end{figure*}

The \CeiO(2--1) emission being optically thin the channel map primarily detects emission from the higher column density molecular cloud associated with G6.55-0.1  (Fig.\,\ref{fig_ceiochanmap}). The primary emission feature in \CeiO(2--1) extended approximately in the north-south direction is due to gas at velocities between 12.5--16\,\kms, while the hub which also corresponds to the maximum intensity shows emission over the entire velocity range.  The \CeiO(2--1) spectrum close to the center of the hub at an offset of (30\arcsec,40\arcsec) show that the emission from this position is due to at least three velocity components (Fig.\,\ref{fig_cospec}).  The north-south extended feature is almost not seen in the channel map for the CO(2--1) emission due to self-absorption (Fig.\,\ref{fig_cochanmap}), but is seen in  \thCO(2--1) with additional diffuse emission (Fig.\,\ref{fig_thcochanmap}). The east-west extended emission feature lying to the north of the CO(2--1) map arises at smaller velocities, consistent with the molecular material associated and physically interacting with the W28 SNR.  The \thCO(2--1)  channel map detects the north-south feature clearly, but shows that while the peak of the emission around 13\,\kms\ lies close to the position of the hub at 14--16\,\kms\ the emission peaks at a position about 50\arcsec\ to the south of the hub, but still lying on the main feature. At velocities beyond 16.5\,\kms\ the north-south emission feature in \CeiO(2--1) appears to be extended toward the north-west going beyond the hub and in the \thCO(2--1) it is connected to a feature around 18.5\,\kms\ towards the south. The \thCO(2--1) emission at velocities exceeding 17\,\kms\ is distinctly fainter than the emission from the molecular gas at 13--15\,\kms. This could explain the lack of connection between the features seen at velocities longer than 17\,\kms\ in the \CeiO (2--1) channel map. Additionally, the  feature seen to the northeast at velocities $<$ 13\,\kms\ though seen as detached peaks in \CeiO(2--1) lie on an extended filament-like structure in \thCO(2--1) which appears to be connected with the north-south feature at the top.

\begin{figure*}
    \centering
    \includegraphics[width = 0.8\textwidth]{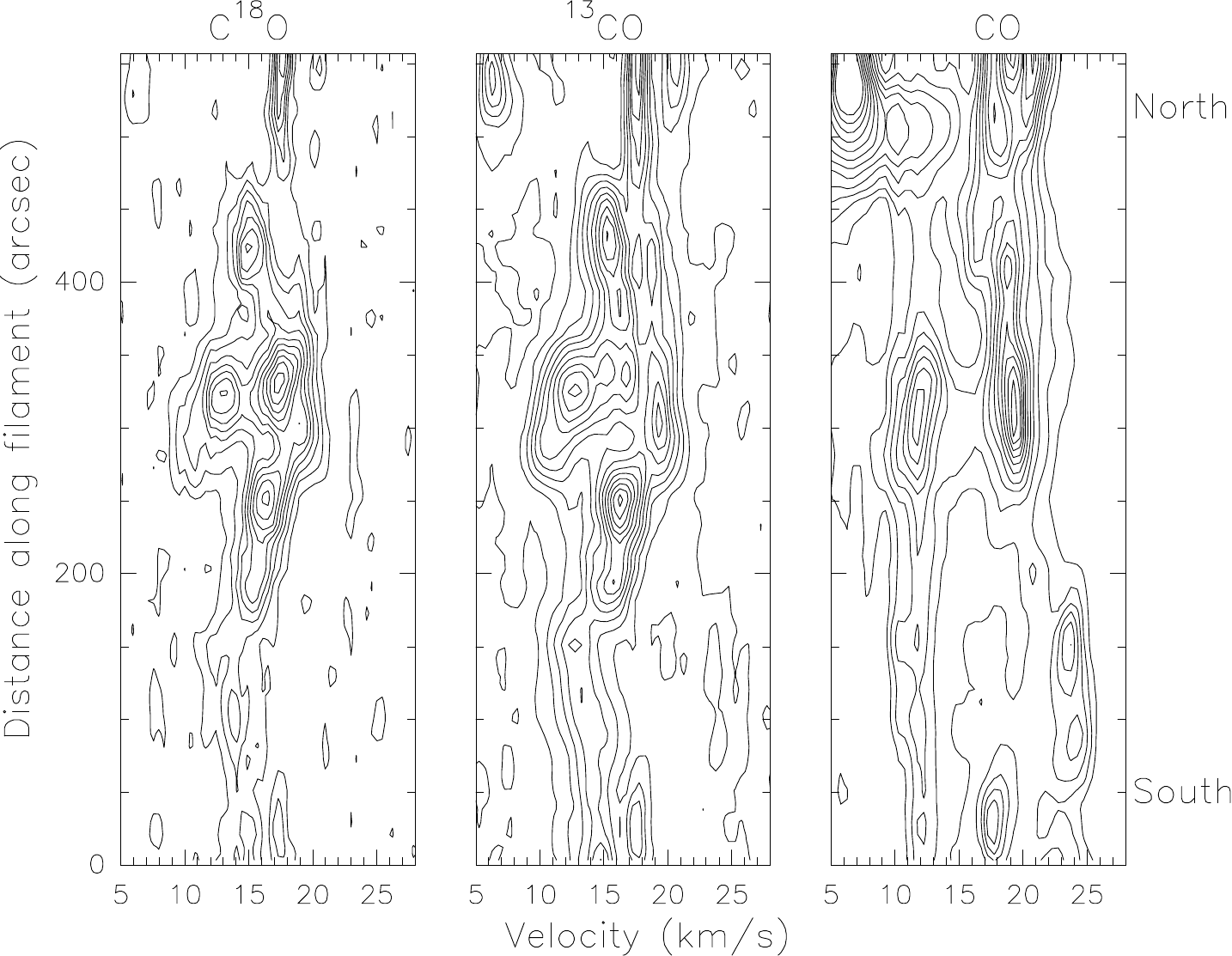}
    \caption{Position-velocity diagrams for \CeiO, \thCO\ and CO (2--1) along the direction shown in Fig.\,\ref{fig_coint}. The contour levels (in K) for \CeiO, \thCO, and CO respectively, are 0.1 to 2.8 (in s.pdf of 0.2), 0.1 to 5.5 (in s.pdf of 0.5) and 1 to 12 (in s.pdf of 1). 
    \label{fig_pvdiags}}
\end{figure*}

In the context of the star formation scenario it is important to understand how and whether the emission features  in the velocity ranges, i.e, $<13$, 13--15\,\kms\, 16--17\,\kms\ and 17--19\,\kms\ in the CO channel maps are kinematically interacting and indeed are the filaments feeding the hub harboring G6.55-0.1.  We have explored the position-velocity diagrams (Fig.\,\ref{fig_pvdiags}) along the direction shown in Fig.\,\ref{fig_coint}. The CO(2--1) $p$-v diagrams possibly quite affected by self-absorption show two narrow features centered around 12 and 20\,\kms. The \thCO\ and \CeiO\ $p$-v diagrams on the other hand show the 16\,\kms\ emission feature to the south of the main emission feature around 13--14\,\kms\ coinciding with the hub in the HFS associated with G6.55-0.1. Exactly at the location of the hub, in both the \CeiO\ and the \thCO\ $p$-v diagrams emission from the 17\,\kms\ cloud is also visible, although only in CO and \CeiO\ it appears to show a ``bridge-like" feature. However, it is not obvious that the two components are indeed interacting with each other or that the mass in the hub is being accreted through both of these components.

\subsection{Gaussian Decomposition of Spectra \label{sec_gausscomp}}

We have used the FUNStools.Decompose \footnote{\url{https://github.com/radioshiny/funstools}} to decompose the \CeiO(2--1) data cubes into multiple Gaussian components. The tool is based on an algorithm that primarily decides the number of components and their velocity positions in the spectrum using the ﬁrst, second, and third derivatives based on the  idea that the velocity component in a ﬁlament is continuous with the surroundings. The results of the first fitting are given as the initial guess to a subsequent round of Gaussian ﬁtting in an iterative manner. The outcome are the parameters such as the central velocity and the linewidth for each velocity component identified at each pixel in the \CeiO\ datacube. We identify coherent structures corresponding to each of the velocity components so identified by applying the friends-of-friends (FoF) technique to the central velocity of each of the components. The algorithm works in an iterative manner in which it first selects the decomposed Gaussian seed component with the maximum amplitude and gives the structure a number. Subsequently the other Gaussian components in the neighboring pixels whose pixel distance is less than $\sqrt{2}$ from the seed component are selected, and the velocity differences of the seed and other components in the neighboring pixels are checked. If the the velocity difference between the seed and other components in the neighboring pixel is less than the velocity dispersion of the seed ($\sigma_\nu^{\rm seed}$) the neighboring component is identified as a friend of the seed and assigned the same group number.  If more than one velocity component in a neighboring pixel is within the range of velocity dispersion from the velocity of the seed, then only the closest one becomes the friend of the seed. In the next iteration the assigned friends of the seed become the seeds of the structure and the same procedure is repeated until there are no more friends to assign.

\begin{figure*}
    \centering
     \includegraphics[width = 0.75\textwidth]{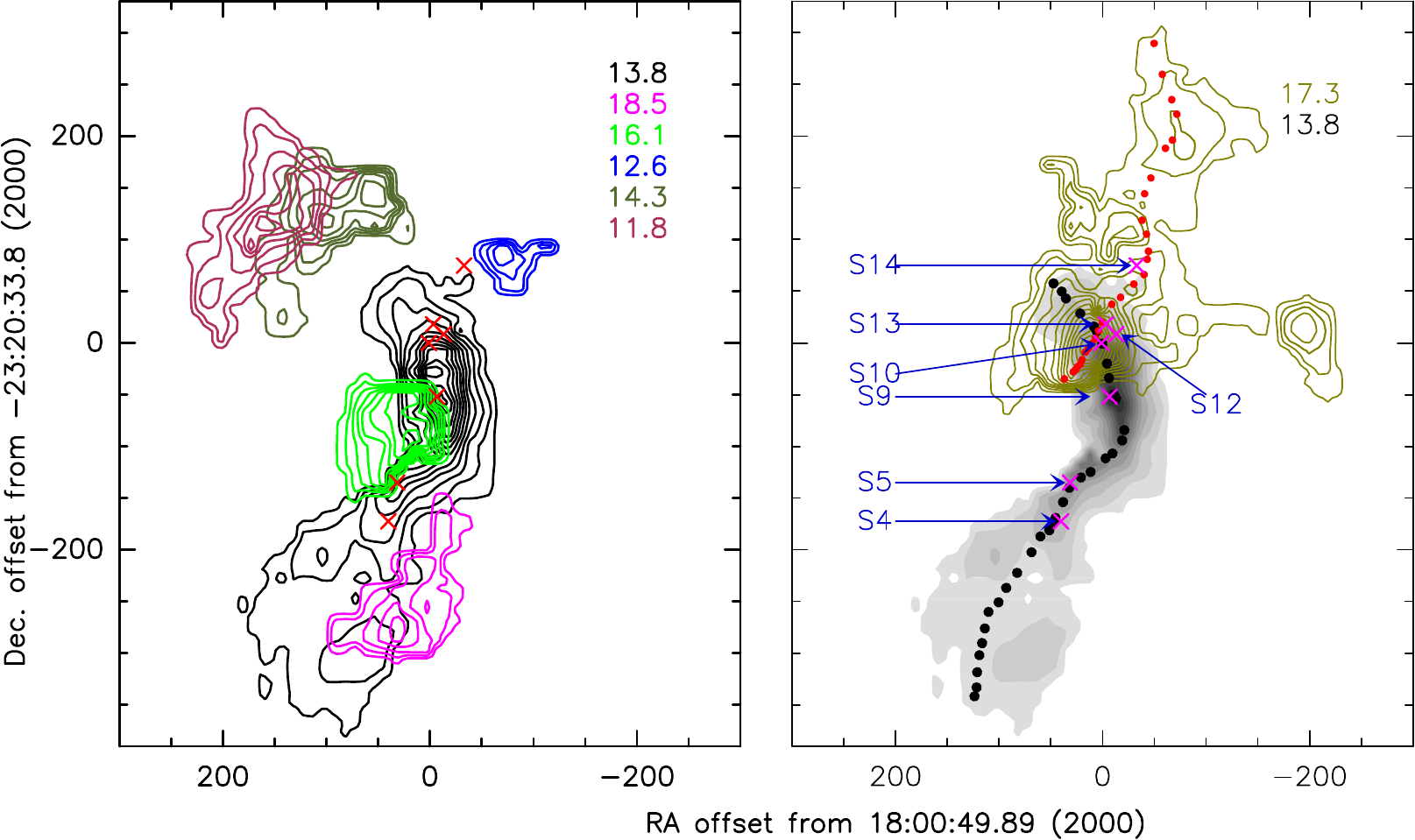}
   \caption{Velocity coherent structures identified in the region based on the \CeiO(2--1) spectra using Gaussian decomposition followed by the identification of groups using a friend-of-friend algorithm. The positions on the 13.8 and 17.3\,\kms\ filaments used to extract the velocity and intensity profiles (Fig.\,\ref{fig_mass_acc_curve}) are marked in black and red respectively, in the right panel. The positions of selected Hi-Gal sources are also marked in the right panel for reference.
    \label{fig_ceiogrp}}
\end{figure*}

Based on the analysis of the \CeiO(2--1) data we identify a total of seven groups centered at velocities (in \kms) of 11.8$\pm$0.4, 12.6$\pm$0.4, 13.8$\pm$0.8, 14.3$\pm$0.4, 16.1$\pm$0.7, 17.3$\pm$1.2 and 18.5$\pm$0.2. Of these, the group centered at 13.8\,\kms\ is the largest coherent structure identified in this region (Fig.\,\ref{fig_ceiogrp}) and the 17.3\,\kms\ component is also reasonably large with emission from the north-western part being quite faint. The 18.5\,\kms\ feature to the south, which appears to be connected to the 17\,\kms\ component in the \thCO(2--1) channel map (Fig.\,\ref{fig_thcochanmap}) is detected separately in the \CeiO\ data. While the groups are consistent with the features seen in the channel maps, exact Gaussian decomposition of the spectra into these components allows for a better quantitative estimate of the properties of the main components.

\subsubsection{Velocity gradient and mass accretion rates along the filaments}

\begin{figure*}
    \centering
    \includegraphics[width = 0.48\textwidth]{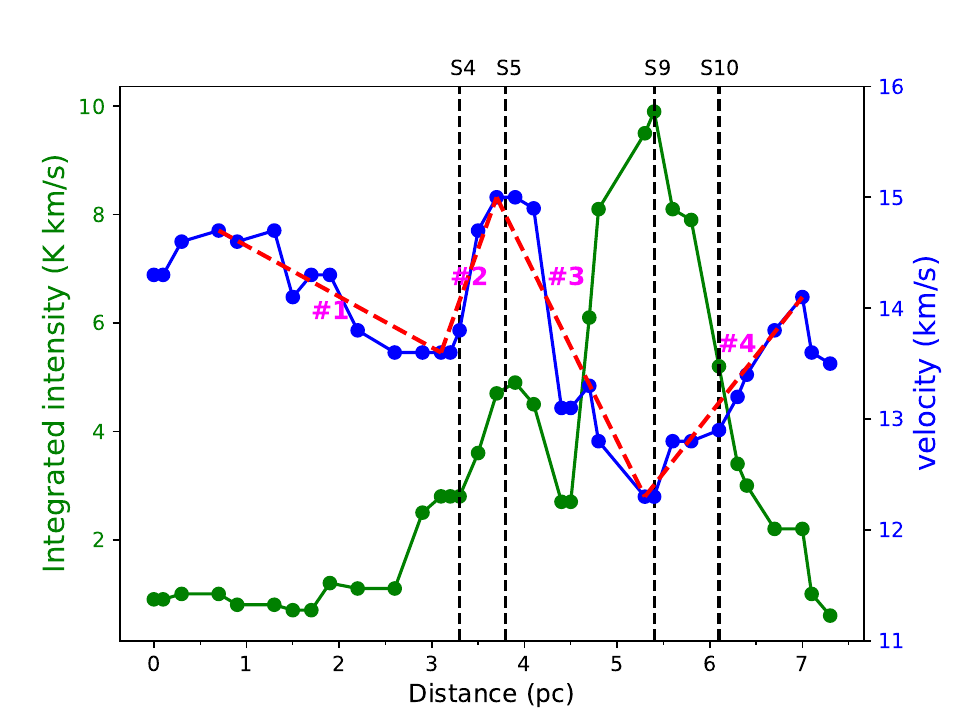} 
    \includegraphics[width = 0.48\textwidth]{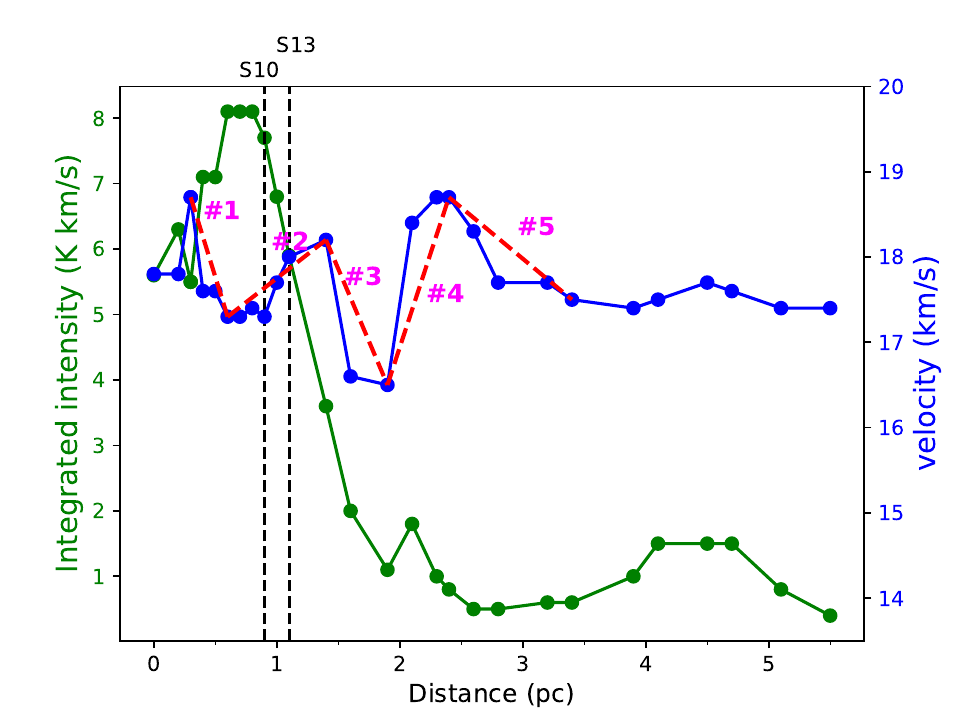}     
    \caption{Variation of velocity and intensity along the length of the velocity features  at 13.8 ({\em Left}) and 17.3\,\kms\ ({\em Right}). Multiple sections of the filament are marked to facilitate discussion in the text. The vertical dashed line shows the positions of the Hi-Gal sources and molecular core identified in the region. 
    \label{fig_mass_acc_curve}}
\end{figure*}

For clouds in gravitational collapse, the gas flow is expected to be accelerated towards the center of the potential well because of gravitational attraction \citep{Gomez2014}. It is expected to produce a characteristic V-shape on the gas velocity structure, which traces the accelerated gas motion around the central high mass clumps or cores. Such features have been detected in other regions including the OMC-1 cloud \citep{Hacar2017} and G333 \citep{Zhou2023}. Figure\,\ref{fig_mass_acc_curve} shows a plot of the velocity and intensity variation along the filaments at 13.8 and 17.3\,\kms. For the ease of discussion and quantitative evaluation we have split up the filaments at 13.8 and 17.3\,\kms\ into four and five segments respectively. 

For the 13.8\,\kms\ filament we find a very strong V-shaped variation of the velocity close to 4.8\,pc from the southern tip of the filament between the segments \#3 and \#4. The tip of the V in the velocity structure aligns exactly with the peak of the intensity profile corresponding to the continuum sources S9 and S10. We note here although S10 is the bright continuum source, the 13.8\,\kms\ feature peaks at the location of S9. Although S9 and S10 being separated by 0.6\,pc are spatially resolved  we attribute the change in velocity gradient to a somewhat larger scale accretion onto both S9 and S10. There is a shallower V-shaped structure in the velocity profile that corresponds to the far-infrared continuum source S4 located right to the south of the hub. Thus we clearly detect signature of mass accretion from the 13.8\,\kms\ filament onto three star forming cores that are also detected in the far-infrared continuum. We have estimated the properties of molecular material in each of the segments assumed for both the filaments (Table\,\ref{tab_velo_mass_accretion}). We calculated the mass accretion rate in each segment of the filament using $\dot{M} = v_{\rm grad} M/{\tan{\theta}}$ \citep{kirk2013, Ma2023} where $M$ is the mass of the segment and $v_{\rm grad}$ is the velocity gradient across it, and  $\theta$ is the angle made by the filamentary part with the plane of the sky which is assumed to be 45\arcdeg. The four segments of the 13.8\,\kms\ filament have masses ranging from 174 to 660\,\msun\ and show mass accretion rates between 152--1020\,\msun\,Myr$^{-1}$. The hub is located between the segments 3 and 4 and shows a total mass accretion rate of $\sim$ 1780\,\msun\,Myr$^{-1}$ from the 13.8\,\kms\ filament. These mass accretion rates are comparable to the values observed in other filamentary star forming systems \citep{Yuan2018,Ma2023,Hu2021}. We have also performed the same analysis for the 17.3\,\kms\ filament and find two V-shaped structures in its velocity profile (Fig.\,\ref{fig_mass_acc_curve}), one at the location of the hub between the segments \#1 and \#2 and the other around 1.5--2\,pc to the north-west from the hub. The second V-shaped structure does not exactly align with a smaller peak in \CeiO(2--1) intensity that seen slightly to the north of the V. The mass accretion rate to the hub from the 17.3\,\kms\ filament is seen to be $\sim 1240$\,\msun\,Myr$^{-1}$, resulting in a total accretion rate of $\sim 3000$\,\msun\,Myr$^{-1}$ from the two filaments.

\begin{deluxetable*}{ccccccccc}
\tablecaption{Properties of segments of the velocity features identified around 13.8 and 17.3\,\kms\ (Fig.\,\ref{fig_mass_acc_curve})
\label{tab_velo_mass_accretion}}
\tablehead{\colhead{Component} & \colhead{Segment} & \colhead{Length} & \colhead{Mass} & \colhead{$v_{\rm grad}$} & \colhead{Mass/Length} & \colhead{$\dot{\rm M}$} & $\Delta V_{\rm deconv}^{\rm med}$ & $M_{\rm crit}^{\rm line}$\\
\nocolhead{} & \nocolhead{} & \colhead{pc} & \colhead{\msun} & \colhead{km\,s$^{-1}$\,pc$^{-1}$} & \colhead{\msun pc$^{-1}$} & \colhead{\msun Myr$^{-1}$} & \colhead{\kms} & \colhead{\msun pc$^{-1}$}}
\startdata
13.8 
& \#1 & 2.4 & 324 & -0.5 & 135& -152 & 2.2 & 490\\
& \#2 & 0.6 & 174 & 2.4 & 290& 415 & 2.1 & 483 \\
& \#3 & 1.6 & 592 & -1.7 & 370 & -1020 & 2.6 & 715\\
&\#4 & 1.7 & 660 & 1.1 & 388 & 715 & 2.6 & 707 \\
\hline
17.3 
&\#1 & 0.3 & 117 & -4.8 & 390 &  -558 & 2.6 & 714\\
&\#2 & 0.8 & 590 & 1.2 & 738 & 679 & 2.4 & 602\\
&\#3 & 0.5 & 109 & -3.5 & 218 & -379 & 1.6 & 296\\
&\#4 & 0.5 & 97 & 4.5 & 194 & 436 & 1.7 & 320\\
&\#5 & 1.0 & 179 & -1.2 & 179 & -220 & 1.1 & 158\\
\enddata
\end{deluxetable*}

\section{Discussion}

\subsection{Is the HFS G6.55-0.1 interacting with the W28 SNR?}

The molecular line emission arising from the north-south extended HFS 6.55-0.1 lies between the velocties of 11 to 18\,\kms, which kinematically corresponds to a distance of 3\,kpc \citep{Urquhart2018}. This suggests that G6.55-0.1 lies well behind the W28 SNR along the line of sight, with the latter being at a distance of 1.9\,kpc. Nevertheless, because of their proximity on sky we carefully searched in our data for spectroscopic evidence of interaction. The SNR W28 has extensive evidence for interaction with molecular clouds, particularly in the form of OH 1720\,MHz maser emission \citep{Claussen1997} and broad molecular line (BML) CO emission \citep{arikawa1999,reach2005}, however these emission were observed from regions that do not overlap with the  HFS G6.55-0.1. The CO(3--2) and (2--1) spectra from the BML regions are clearly marked by line widths of $\sim 20$ to 30\,\kms, with maximum values of up to 70\,\kms\ and sharply peaked profiles that are typical of shocked gas \citep{reach2005}. The spectral line observed in the main body of the HFS 6.55-0.1 for all the three tracers typically have much smaller widths and do not have shapes that are indicative of shocked gas. The east-west extended feature lying to the north of the HFS is detected only in CO(2--1) with the main emission centered at velocities less than 7\,\kms\ and with widths $>$5\,\kms, which is in contrast to the velocity components identified in the HFS. Although the CO(2--1) emission is strongly affected by opacity issues, a clear discontinuity in emission between  even 9--11\,\kms\ can be seen in the channel map itself (Fig.\,\ref{fig_cochanmap}). Further neither of \CeiO\ and \thCO\ (2--1) spectra show any signs of being affected by any external forces particularly towards the north and the east. We thus conclude, that HFS G6.55-0.1 shows no evidence for physical interaction with the W28 SNR and hence the star formation in G6.55-0.1 is unlikely to be triggered by the impact of the SNR.

\subsection{Gravitational Stability of the filaments \& cores in G6.55-0.1}

The velocity distribution of the molecular material in the region around G6.55-0.1 being highly complex, Gaussian decomposition of the spectra was necessary to measure the typical linewidth corresponding to each of the velocity components. Due to the large velocity gradients, particularly at the location of the hub, part of the observed linewidth is due to the shift in velocity over the beamsize of 26\farcs2 of the \CeiO\ data. We have corrected for the effect of beam-broadening on the line width by considering the velocities of all pixels within 27\arcsec\ of a particular pixel. Following the method described by \citet{Stil2002} we estimate the beam-deconvolved linewidth from the observed linewidth using $\sigma_{\rm deconv}^2 = \sigma_{\rm obs}^2 - \frac{1}{2}\left(V_{\rm max} - V_{\rm min}\right)^2$, where $V_{\rm max}$ and $V_{\rm min}$ are the maximum and minimum values of $V_{\rm cen}$ found within a beam centered at the particular pixel and the $\sigma$s are related to the observed and deconvolved linewidths through $\Delta V = \sqrt{8\ln{2}}\sigma$. We have assumed a constant intensity over the beamwidth resulting in the factor of 1/2 in the relation between observed and deconvolved $\sigma$s The median deconvolved linewidth of each segment of the two filaments at 13.8 and 17.3\,\kms\ are presented in Table\,\ref{tab_velo_mass_accretion}. The deconvolved linewidth so obtained is a combination of contributions from both thermal and non-thermal gas motions. To estimate the non-thermal motions associated with gas turbulence or core formation motions we have subtracted the thermal component from the beam-deconvolved linewidth assuming that the two contributions are independent of each other so they add in quadrature (e.g., Myers 1983). We estimate the non-thermal velocity dispersion as follows:

 \begin{equation}
 \sigma_{\rm NT} = \displaystyle\sqrt{\frac{\Delta {\rm V_{\rm deconv}^2}}{8\ln 2}-\frac{kT}{m_{\rm mol}}}   
\end{equation}

where $k$ the Boltzmann’s constant, T the gas kinetic temperature, and $m_{\rm mol}$ the mass of the molecule under consideration. The velocity dispersion $\sigma_{\rm NT}$ can be directly compared to the (isothermal) sound speed of the gas, $c_s$, which has a value of 0.27\,\kms\ for ISM gas at 20 K. We use the median values of beam-deconvolved linewidths for each of the segments identified in the two filaments and estimate the critical line mass of each segment for a temperature of 20\,K following \citep{Andre2014}

\begin{equation}
 M^{\rm line}_{\rm crit} = \left[1+\left(\frac{\sigma_{\rm NT}}{c_s}\right)^2\right]\left[16{\rm M_{\odot}pc}^{-1}\times \left(\frac{T}{{\rm 10 K}}\right)\right]   
\end{equation}

 For all the segments  of the 13.8 and 17.3\,\kms\ filaments the mass per unit length estimated for the different parts of the two filaments are consistently smaller than the critical line masses derived for the segments (Table\,\ref{tab_velo_mass_accretion}). This suggests that although the filaments show clear kinematic signature of mass accretion, as a whole these are not in a state of gravitational collapse. Considering typical velocity widths of 2.4\,\kms\ the critical line mass at 20K is estimated to be 498.1\,\msun pc$^{-1}$, which is smaller than the mass per unit length of 594\,\msun pc$^{-1}$ observed for the entire filament with a mass of 4520\,\msun\ and a length of 7.6\,pc. In contrast to the values presented in Table\,\ref{tab_velo_mass_accretion} here we have considered the total mass of the HFS estimated over all the velocity components associated with it and comparison of this estimate with the critical line mass  suggests that the HFS as a whole is likely to be self-gravitating.

We estimate the gravitational stability of the 16.1\,\kms\ cloud by calculating the virial parameter defined as $\alpha_{\rm vir} = 1.2\left(\frac{\Delta V}{\rm km\,s^{-1}}\right)^2 \left(\frac{R}{\rm pc}\right) \left(\frac{\rm M}{10^3\,\msun}\right)^{-1}$, such that for values of $\alpha \sim <1$ the cloud is gravitationally unstable. For the 16.1\,\kms\ cloud we use an effective radius of  0.8\,pc, a total mass of 610\,\msun\ and a beam-deconvolved linewidth of 1.5\,\kms\ to obtain $\alpha_{\rm vir} \sim 8$. This suggests that the cloud at 16.1\,\kms\ is supported against gravitational collapse by turbulent gas motions. 

We note that due to ongoing high-mass star formation in the hub the gas temperature of the 16.1\,\kms\ cloud as well as the segments of the 13.8 ad 17.3\,\kms\ filaments which are close to the hub are likely to be significantly higher than the assumed value. This could lead to somewhat lower estimates for the non-thermal velocity component.

The velocity gradient arising due to free-fall under gravity is related to the mass and size of the collapsing region by $\upsilon_{\rm grad} = \displaystyle\sqrt\frac{GM}{2R^3}$. From the H$_2$ column density estimated from the \CeiO(2--1) emission we estimate the mass of the hub within a radius of 30\arcsec\ (equal to 0.45\,pc) in G6.55-0.1 to be 554\,\msun. The velocity gradient corresponding to the free-frall of such a clump is estimated to be 3.6\,\kms\,pc$^{-1}$, a value which is consistent with the range of velocity gradients seen in the segments of the 13.8 and 17.3\,\kms\ filaments that are close to the hub. We note that the velocity gradients derived from the observations are subject to uncertainties due to the assumed inclination of the filament with the plane of the sky. Thus, the velocity gradients obtained around in the hub can be consistently explained in terms of gravitational collapse. 

\subsection{Is G6.55-0.1 really a HFS?}

The source  G6.55-0.1 was identified as a HFS based on dust continuum observations. We have used spectroscopic observations to identify the kinematically related structures (filaments) merging at the location of the associated \HII\ region (hub). We find that the emission from molecular gas arises due to multiple velocity components three of which at 13.8, 16.1 and 17.3\,\kms\ appear to be directly associated with the source G6.55-0.1. Of these, the components at 13.8 and 17.3\,\kms\ appear to overlap at the hub, while the 16.1\,\kms\ is seen a little to the south of the hub. The 13.8\,\kms\ component shows two clear signatures of accelerated gas flow due to mass accretion leading to the formation of the young stellar objects S9, S10 (G6.55-0.1) and S4. The 17.3\,\kms\ component emission is strongly peaked at the location of S10 in the hub, while for the 13.8\,\kms\ although the intensity is enhanced in the hub, the absolute peak is slightly to the south of the hub coinciding with the source S9. Both the 13.8 and 17.3\,\kms\ features however show a strong change in the velocity gradient at the location of the hub (Fig.\,\ref{fig_mass_acc_curve}). In order to ensure that the velocity gradients so derived for the filaments intersecting at the position of the hub do not suffer from any artefact due to the automated fitting procedure used in this work , we have checked the gaussian fits within the hub manually as well, an example of which is shown in Fig.\,\ref{fig_hubfit}. At the position of the center of the hub which corresponds to an ATLASGAL source, NH$_3$ observations with a resolution of 40\arcsec\ have identified two components at 12.8 and 16.1\,\kms. This could be indicative of the fact that the 17.3\,\kms\ component does not arise from gas at densities high enough to excite the NH$_3$ lines. For the 17.3\,\kms\ component based on the mass derived for the region around the peak we estimate an average density of 900\,\cmcub. 

The \CeiO (2--1) spectra observed in the hub (Fig.\,\ref{fig_hubfit}) are qualitatively similar to the profiles predicted by smooth particle hydrodynamic simulations of a Jeans-unstable dense prolate clump in the process of collapsing along its long axis on a near free-fall dynamical timescale \citep{Peretto2007,Peretto2006}. For NGC 2264-C for a central core of 90\,\msun\ the velocity difference between the two peaks is 2\,\kms, which is possibly consistent with the 3.6\,\kms\ velocity gap that we observe for  G6.55-0.1 where we do not resolve the central core but can only observe a bigger clump with a mass of 554\,\msun. The V-shaped velocity profiles with locations of the largest gradient coinciding with the intensity peak (Fig.\,\ref{fig_intprof}) is indicative of accretion of mass from both directions. Additionally, \citet{Zhou2023} suggest that global hierarchical collapse under gravity would create a funnel-shaped structure in the position-position-velocity space and the V-shaped velocity structure along the filament skeleton are likely to be a projection of this structure to the position-velocity plane. Though we do not spatially resolve the central accreting core, the spectral profiles, the V-shaped velocity profile and the far-infrared continuum sources together suggest that the hub is likely undergoing collapse. We note that the segment-wise virial analysis of the two main filaments suggest the filaments to have masses smaller than their respective critical line masses (Table\,\ref{tab_velo_mass_accretion}).  This apparent contradiction could arise from an overestimate of $\sigma_{\rm NT}$ due to enhanced gas temperature and turbulence due to the high mass stars already forming in the hub. On the other hand, the outcome of the self-gravitating clouds being  stable on larger scales and the clumps on parsec scales being dynamically decoupled from their surrounding molecular cloud and collapsing has also been observed in several other sources \citep[e.g.,][]{Mookerjea2023,Peretto2023}. We thus conclude that the high mass star formation in G6.55-0.1 occurs in a hub that is being fed by at least two filaments at 13.8 and 17.3\,\kms.

\section{Summary \& Conclusions}

We have explored the morphology and kinematics of the source G6.55-0.1, which is a massive star-forming region associated with a hub-filament system. The newly obtained radio continuum maps detect multiple peaks that are associated with far-infrared sources, thus confirming the presence of multiple high mass protostars in the hub region. The total radio continuum flux detected from the \HII\ region can be explained by a single ZAMS O8.5V star or multiple B type stars.  We used the velocity information available from the $J$=2--1 transitions of CO (and its isotopes) to identify velocity-coherent structures (filaments and cores) and investigate signatures of mass accretion along the filaments to the hubs. The molecular emission associated with G6.55-0.1 is strongly affected by the velocity crowding due to multiple overlapping components along the line of sight (including the W28 SNR) as well as by the optical thickness of both CO and \thCO(2--1) lines. The \CeiO(2--1) line though optically thin being fainter primarily traces the higher column density filaments and the hub. However, comparison of the CO and \CeiO\ spectra, particularly towards the north-east where the cloud overlaps with molecular gas affected by the W28 SNR, enabled us to confirm that G6.55-0.1 is physically not associated with W28. The column density of molecular gas in the region derived from both dust continuum emission and \CeiO (2--1) intensities were found to be consistent within a factor of 2 with a median value of 7$\times$10$^{22}$\,\cmsq, a value typical of similar HFSs in which massive star formation is seen \citep{Zhou2022}. An automated Gaussian decomposition followed by identification of correlated structures using a Friend-of-Friend algorithm was used to disentangle the complex \CeiO(2--1) spectra. The components segregated using this procedure confirmed the interaction of only two of the filaments with median velocities of 13.8\,\kms\ and 17.3\,\kms\ at the location of the hub. The other velocity-coherent structures are either more core-like or even if these are extended like the 11.8\,\kms\ feature the connection to the hub is tenuous possibly because the \CeiO(2--1) emission is too faint to be detected. Use of \thCO\ and CO spectra to confirm any interaction is unreliable due to the contamination by the velocity components physically not associated with the source as well as due to the optical thickness of the emission resulting in self-absorption. Both the filaments at 13.8 and 17.3\,\kms\ show increased velocity gradients and V-shaped structure in the hub that harbors the \HII\ region that suggest gravitational collapse with mass accretion rates between 558--1020\,\msun\,Myr$^{-1}$. The mass accretion rates typically found in star forming regions range from a few 10 to a few 100\,\msun\,Myr$^{-1}$ \citep[][and references therein]{Hacar2022}, with the high-mass star-forming HFSs showing up to 1500\,\msun\,Myr$^{-1}$ \citep{Ma2023,Hu2021}. The combined accretion rate towards the hub is 3000\,\msun\,Myr$^{-1}$, which though on the higher side is consistent with the previous observations within the uncertainties of the inclination of the filaments on the plane of the sky. Segment-wise analysis of the filaments based on virial parameters suggest the filament to be stable, while the gas in the hub shows spectral and velocity profiles indicative of dynamical decoupling/collapse leading to the formation of high mass stars. It is also likely that the filaments being identified at a resolution of 0.4\,pc have coherent sub-structures that will require a similar analysis.  Follow-up higher resolution observations of high density tracers, preferably without hyperfine structures, are needed to constrain the properties of the filaments and the flow of material through them.

\appendix

\section{Extra figures}

\begin{figure*}
    \centering
    \includegraphics[width = 0.7\textwidth]{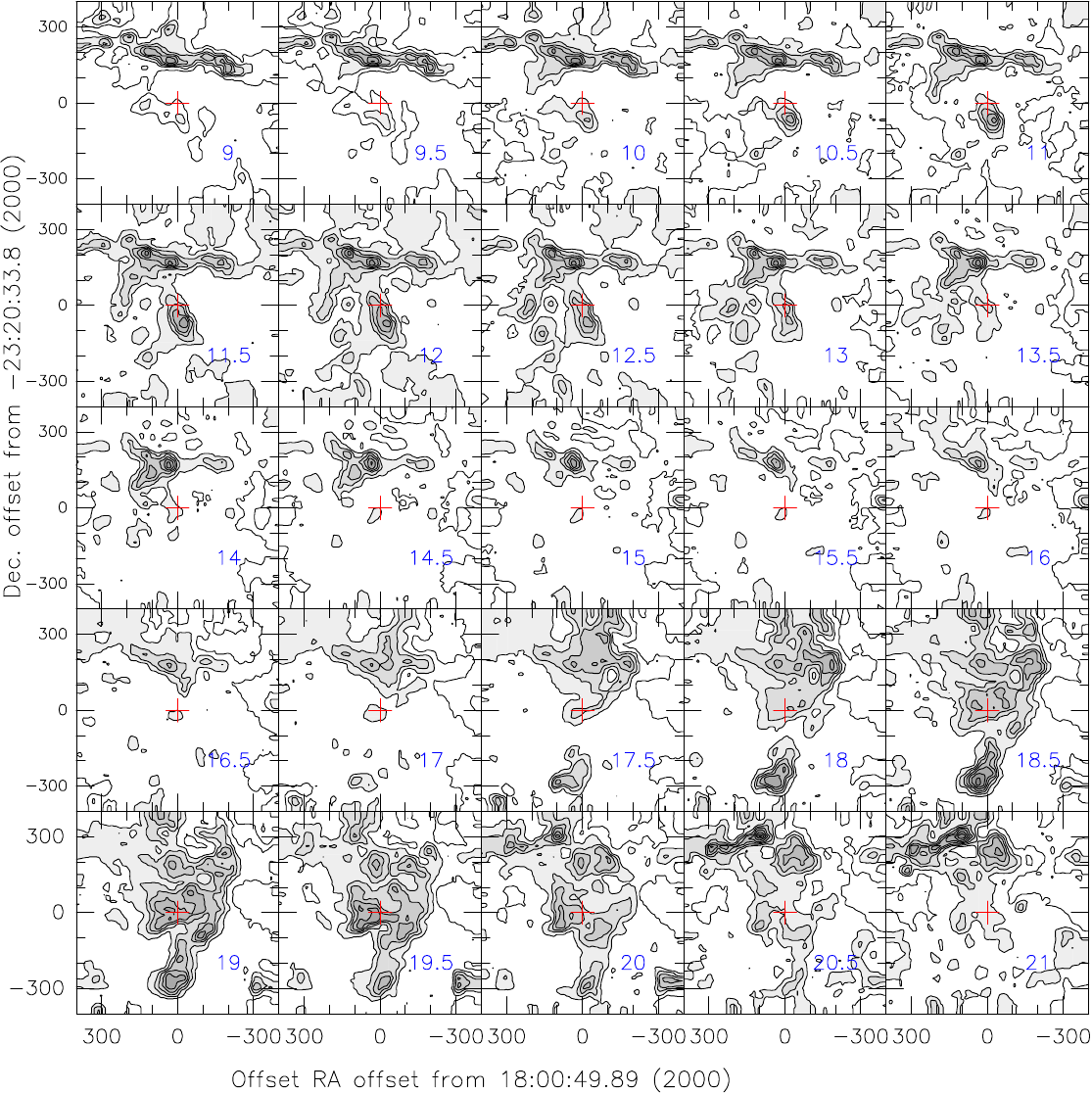}
    \caption{Velocity-channel map of the CO(2--1) emission. The contours are at 1 to 19\,\kms\ in s.pdf of 2\,\kms. The rest of the details are the same as in Fig.\,\ref{fig_cochanmap}.
    \label{fig_cochanmap}}
\end{figure*}

\begin{figure*}
    \centering
    \includegraphics[width = 0.7\textwidth]{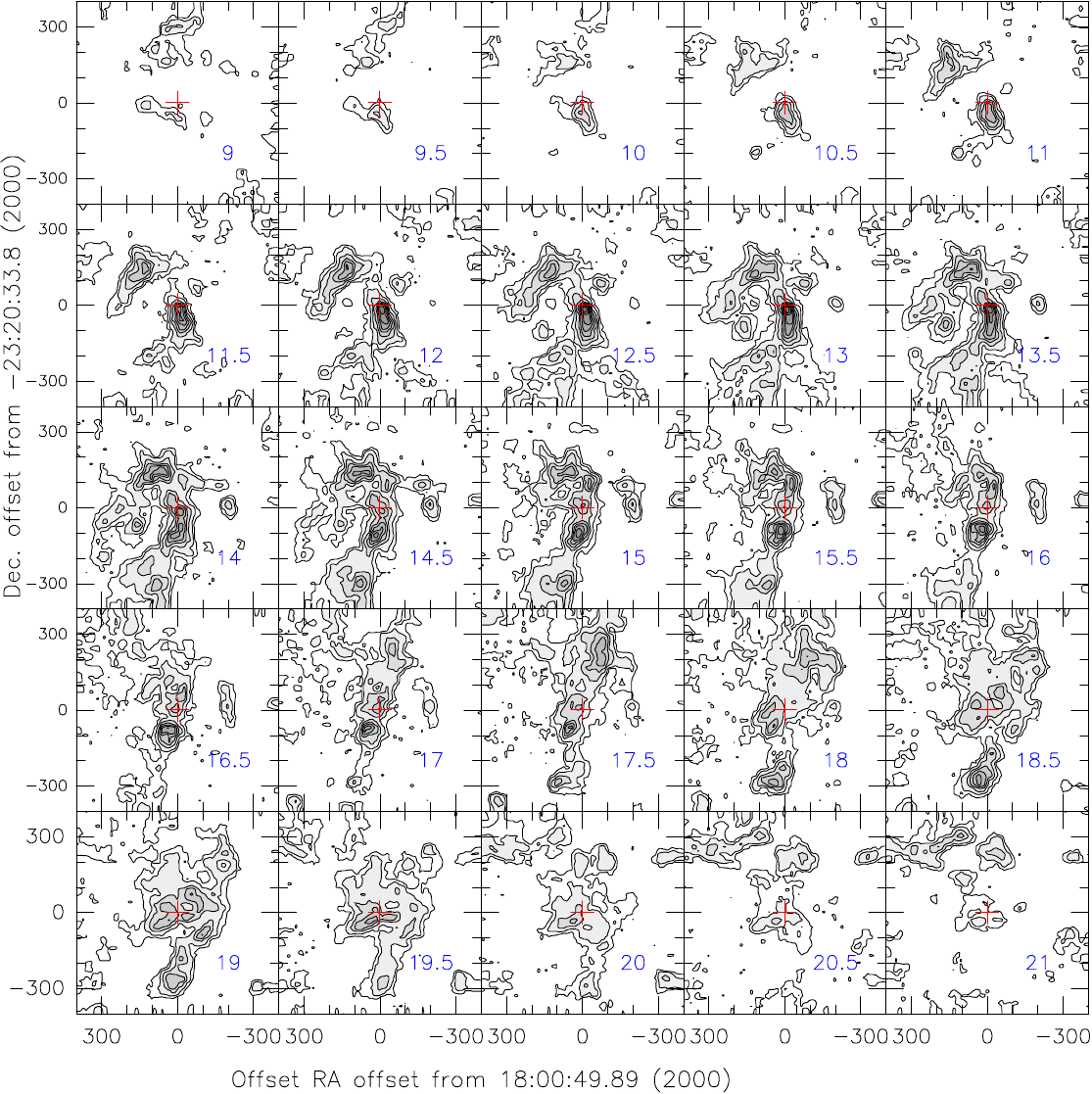}
    \caption{Velocity-channel map of the \thCO(2--1) emission. The contours are at 0.5 to 8.5\,\kms\ in s.pdf of 1\,\kms.
    \label{fig_thcochanmap}}
\end{figure*}

\begin{figure}
\centering
\includegraphics[width = 0.45\textwidth]{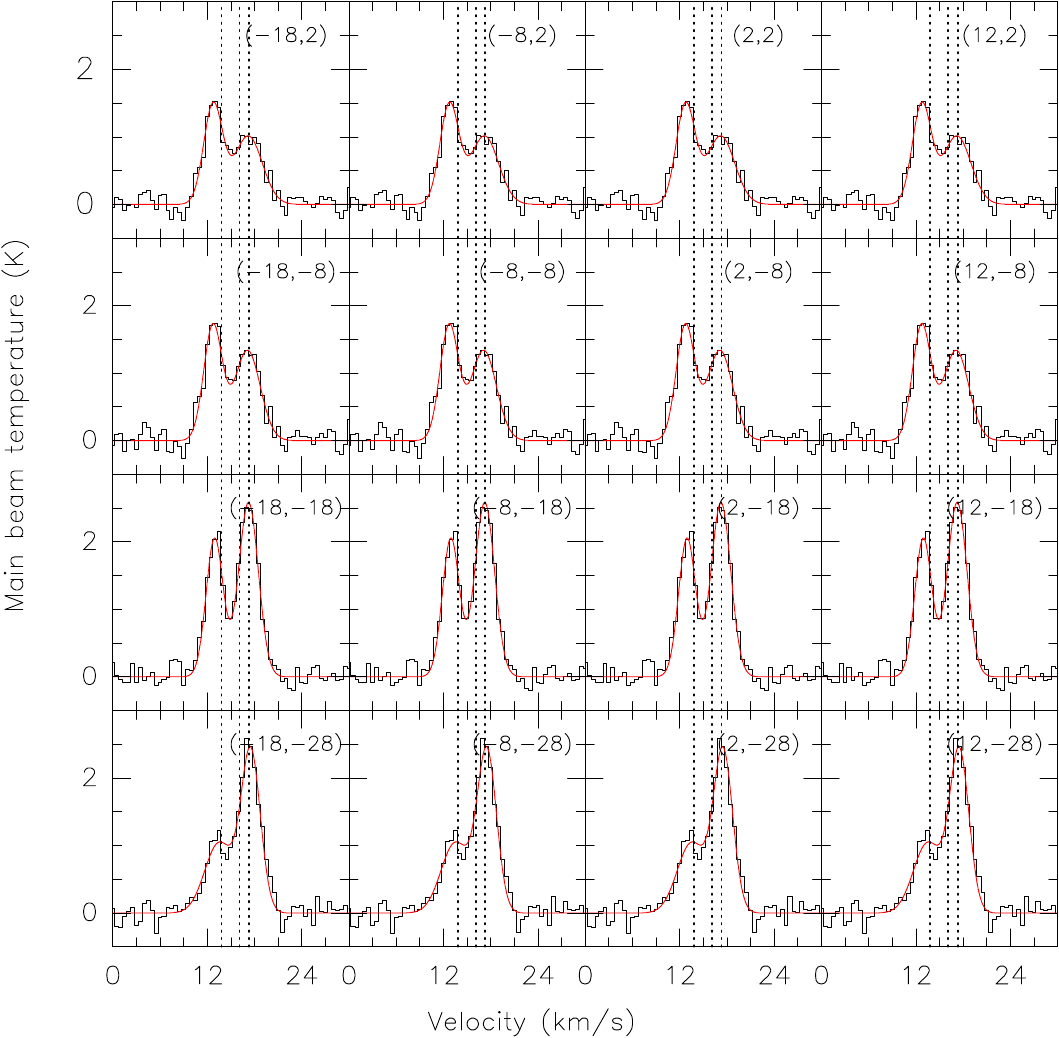}
\caption{Multi-component Gaussian fits (in red) to \CeiO(2--1) spectra (in black) at positions around the center of the hub in G6.55-0.1. The dashed vertical lines are drawn at 13.8, 16.1 and 17.3\,\kms\ to guide the eye.}
\label{fig_hubfit}
\end{figure}

\begin{acknowledgements}
Authors thank the referee for comments and suggestions that helped improve the paper significantly. S. Sen and B. Mookerjea acknowledge the support of the Department of Atomic Energy, Government of India, under Project Identification No. RTI 4002. C. H. Ishwara-chandra acknowledges the Department of Atomic Energy for funding support, under
project 12-R\&D-TFR-5.02-0700. We thank the staff of the GMRT that made these observations possible. GMRT is run by the National Centre for Radio Astrophysics of the Tata Institute of Fundamental Research.This research has made use of the NASA/IPAC Infrared Science Archive, which is funded by the National Aeronautics and Space Administration and operated by the California Institute of Technology.
\end{acknowledgements}








\end{document}